%% file: egpaper.tex
\documentclass[10pt,twocolumn,letterpaper]{article}

\usepackage{wacv}
\usepackage{times}
\usepackage{graphicx}
\usepackage{epsfig}
\usepackage{amsmath}
\usepackage{amssymb}
\usepackage{booktabs}

\def\wacvPaperID{112} 

\wacvfinalcopy 

\ifwacvfinal
\def\assignedStartPage{1} 
\fi

\ifwacvfinal
\usepackage[breaklinks=true,bookmarks=false]{hyperref}
\else
\usepackage[pagebackref=true,breaklinks=true,colorlinks,bookmarks=false]{hyperref}
\fi

\ifwacvfinal
\else
\fi

\newcommand{\ppg}{\url{https://bvraghav.com/shad3s/}}

\begin{document}

\title{\textsc{shad3s}: A model to Sketch, Shade and Shadow}

\author{
  Raghav Brahmadesam Venkataramaiyer, Abhishek Joshi, Saisha Narang,
  Vinay P. Namboodiri\\
  {\tt\small \{bvraghav,abhishekjoshi,saisha,vinaypn\}@iitk.ac.in} \\
  \textsc{indian institute of technology kanpur}
}

\maketitle

\begin{abstract}
  Hatching is a common method used by artists to accentuate the third
  dimension of a sketch, and to illuminate the scene.  Our system
  \textsc{shad3s}\footnote{\ppg --- The project page; hosted with
    further resources.} attempts to compete with a human at hatching
  generic three-dimensional (\textsc{3d}) shapes, and also tries to
  assist her in a form exploration exercise. The novelty of our
  approach lies in the fact that we make no assumptions about the
  input other than that it represents a \textsc{3d} shape, and yet,
  given a contextual information of illumination and texture, we
  synthesise an accurate hatch pattern over the sketch, without access
  to \textsc{3d} or pseudo \textsc{3d}. In the process, we contribute
  towards \emph{a)} a cheap yet effective method to synthesise a
  sufficiently large high fidelity dataset, pertinent to task;
  \emph{b)} creating a pipeline with conditional generative
  adversarial network (\textsc{cgan}); and \emph{c)} creating an
  interactive utility with \textsc{gimp}, that is a tool for artists
  to engage with automated hatching or a form-exploration exercise.
  User evaluation of the tool suggests that the model performance does
  generalise satisfactorily over diverse input, both in terms of style
  as well as shape. A simple comparison of inception scores suggest
  that the generated distribution is as diverse as the ground truth.
\end{abstract}


\begin{figure*}
  \centering
  \includegraphics[width=\textwidth]{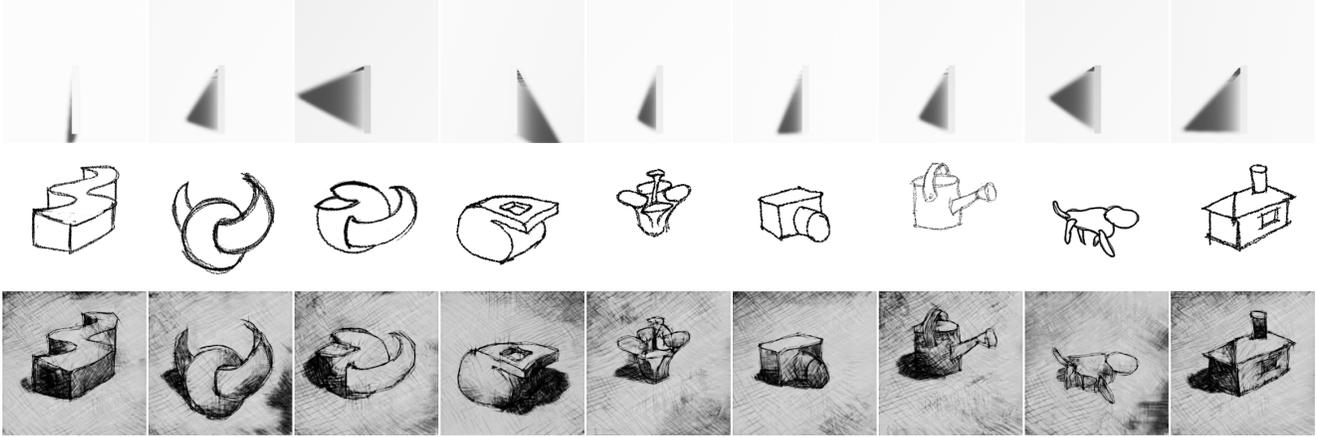}
  \caption{\textsc{shad3s}: a Model for Sketch, Shade and Shadow ---
    is a completion framework that provides realistic hatching for an
    input line drawn sketch that is consistent with the underlying
    \textsc{3d} and a user specified illumination. The figure above
    shows choice of illumination conditions on the top row, user
    sketches in the middle row, and the completion suggestions by our
    system. Note that the wide variety of shapes drawn by the user, as
    well as the wide variations in brush styles used by the user were
    not available in the training set.}
  \label{fig:teaser}
\end{figure*}


\section{Introduction}
\label{sec:introduction}

Sketches are a widely used representation for designing new objects,
visualizing a \textsc{3d} scene, for entertainment and various other
use-cases. There have been a number of tools that have been developed
to ease the workflow for the artists. These include impressive work in
terms of non-photorealistic rendering (\textsc{npr}) that obtain
sketches given \textsc{3d} models
\cite{Hertzmann:2000:ISS:344779.345074,gerl:hal-00781065}, others that
solve for obtaining \textsc{3d} models given sketch as an input using
interaction
\cite{Han:2017:DDL:3072959.3073629,Delanoy:2018:SUM:3242771.3203197},
and several others that aim to ease the animation task
\cite{Benard:2013:SAE:2461912.2461929,Jamriska:2015:LAT:2809654.2766983,Fiser:2017:ESS:3072959.3073660}. However,
these works still rely on the input sketch being fully generated by
the artist. One of the important requirements to obtain a good
realistic sketch is the need to provide the hatching that conveys the
\textsc{3d} and illumination information. This is a time consuming task and
requires effort from the artist to generate the hatching that is
consistent with the \textsc{3d} shape and lighting for each sketch
drawn. Towards easing this task, we propose a method that provides
realistic hatching for an input line drawn sketch that is consistent
with the underlying \textsc{3d} and illumination.

Though recent works have tried to address problems associated with
shading sketch~\cite{zheng_learning_2020}, however, to the best of our study
we couldn't find any prior work addressing the problem we intend to
solve. 
Here we hope to leverage deep learning to decode and translate the
underlying \textsc{3d} information amongst a rasterised set of
\textsc{2d} lines. Deep learning has been promising to solve many
interesting and challenging
problems~\cite{NIPS2014_5423,DBLP:journals/corr/MirzaO14,isola_image--image_2016,Delanoy:2018:SUM:3242771.3203197}. Moreover,
it is well known that deep learning algorithms are data
intensive. There exist
datasets~\cite{Delanoy:2018:SUM:3242771.3203197,zheng_learning_2020}
which aim to solve \textsc{3d}-related problems like predicting the
illumination, and mesh inference from line drawings, but they are
limited in their diversity and level of pertinence (see
\S~\ref{sub-sec:how_stand_out}). This motivates us to contribute the
\textsc{shad3s} dataset for \textit{Sketch-Shade-Shadow} like tasks
(see \S~\ref{sec:dataset}). Hereby, we attempt to help the research
community to bring closer, the two domains of graphics and vision.

With the aim of creating artistic shadows with hatch-patterns for a
given contour drawing of a \textsc{3d} object, under a user specified
illumination, we define this research problem as an attempt \emph{to
  learn a function}, that maps a combination of --- \emph{a)}
hand-drawn \textsc{3d} objects; \emph{b)} illumination conditions; and
\emph{c)} textures --- to a space of completed sketches with shadows
manifested as hatch-patterns. See \S~\ref{sec:methodology}.

Our proposed pipeline is based on training a conditional generative
adversarial network (\textsc{cgan})~\cite{DBLP:journals/corr/MirzaO14}
with a data-set of \textsc{3d} shape primitives and aims to model the
problem as one of contextual scene completion. While this approach has
been widely explored in the image
domain~\cite{long_fully_2015,isola_image--image_2016}, the challenge
is to ensure that we are able to obtain convincing results in sketch
domain using a sparse set of lines. In order to solve this we train
our model using a novel \textsc{shad3s} dataset explained further in
\S~\ref{sec:dataset}. These are rendered to be consistent with the
illumination and \textsc{3d} shape. We include the context required
for solving the problem in the input. Once this is learned it is
possible to automate the sketch generation using the proposed
\textsc{cgan}
model. 
A glimpse of our results can be seen in Fig.~\ref{fig:teaser}, that
once again reinforces the significance of sufficiently large dataset
in the context of deep learning, and its generalizability.

A natural approach given a dataset would be to train a regression
based system that would aim to generate the exact information that is
missing. This can be achieved using reconstruction based losses. The
drawback of such an approach is that the resultant generation would be
blurred as it averages over the multiple outputs that could be
generated. In contrast, our adversarial learning based approach allows
us to generate a sharp realistic generation that we show is remarkably
close to the actual hatch lines.

The system has been tested for usability and deployment under casual
as well as involved circumstances among more than 50 participants, the
qualitative results of which are selectively displayed here (see
\S~\ref{sec:user-evaluation}), and a larger subset has been made
available as supplement. We would like to mention that the samples for
which the system has been tested by the user \emph{are very different}
from the distribution of samples for which it was trained on. It is
able to generalize and produce satisfactory sketches as shown in
Fig. \ref{fig:teaser}. Through this paper we provide, to the best of
our knowledge, the first such tool that can be used by artists to
automate the tedious task of hatching each sketch input. The tool and
the synthetic dataset (described in \S~\ref{sec:dataset}) are
available for public use at the project page\footnote{\ppg}.

\paragraph{Summary.}

To provide an overview of our work, we summarise our main
contributions as follows. \textit{Firstly,} we define an extremely
inexpensive method to generate sufficiently large number of high
fidelity stroke data, and contribute the resultant novel public
dataset.

\textit{Secondly,} we define simple method to inject data from three
different modalities, namely: \textit{i)} sketch representing
\textsc{3d} information; \textit{ii)} rendered canonical geometry
representing illumination; and \textit{iii)} a set of hand drawn
textures representing multiple levels of illumination — these three
may be plugged into any well established deep generative model.

\textit{Finally,} we push boundaries for artists through an
interactive tool to automate a tedious hatching task, and to serve as
a form-exploration assistant.





\section {Relevant Works}
\label{sec:relevant-works}

As the proposed method is the first work to address the task of
automating hatch generation from sketches, there is no directly
related work to the best of our knowledge.  However, there are two
prominent lines of approach, namely stylization and sketch-based
modeling that are related to our problem. In contrast to these
approaches, we aim to address the task as one of contextual
information completion and therefore the proposed approach differs
substantially from these approaches.




\subsection{Stylization}
\label{sec:stylization}

\paragraph{An undercurrent,} is arguably visible in a series of works
following Siggraph '88, where in a panel proceedings
\cite{Mackinlay:1988:DEP:1402242.1402254}, Hagen said,
\begin{quote}
  The goal of effective representational image making, whether you
  paint in oil or in numbers, is to select and manipulate visual
  information in order to direct the viewer's attention and determine
  the viewer's perception.
\end{quote}



We see this in the initial works
\cite{Haeberli:1990:PNA:97880.97902,Salisbury:1994:IPI:192161.192185,Curtis:1997:CW:258734.258896}
focusing on attribute sampling from user input to recreate the image
with arguably a different visual language and accentuation and offer
interactivity as a medium of control. Thereby followed, a shift
towards higher order controls, to achieve finer details, for example
temporal coherence~\cite{Litwinowicz:1997:PIV:258734.258893},
modelling brush strokes~\cite{Hertzmann:1998:PRC:280814.280951}, and
space-filling~\cite{Shiraishi:2000:AAP:340916.340923}. Space-filling
had been buttressed, on one hand with visual-fixation data and
visual-acuity~\cite{DeCarlo:2002:SAP:566654.566650,Santella:2002:APR:508530.508544},
and on the other with semantic units inferred using classical vision
techniques~\cite{Zeng:2009:IPP:1640443.1640445}. The finer levels of
control, in these works, allowed for interactivity with detail in
different parts of the synthesised image.

\paragraph{Image Analogies}

had popularised the machine learning framework for stylising images,
with analogous reasoning $A:A'::B:B'$, to synthesise $B'$, using image
matching~\cite{Hertzmann:2001:IA:383259.383295}; to achieve temporal
coherence~\cite{Benard:2013:SAE:2461912.2461929}; using synthesis over
retrieved best-matching
feature~\cite{Fiser:2017:ESS:3072959.3073660,fiser_stylit_2016,Jamriska:2015:LAT:2809654.2766983}.
The problems thus formulated, allowed use of high level information as
a reference to a style, for example a Picasso's painting.

\paragraph{Deep Learning Techniques}

have recently documented success in similar context, using Conditional
Generative Adversarial Networks (\textsc{cgan})
\cite{DBLP:journals/corr/MirzaO14}, and its variants using sparse
information~\cite{isola_image--image_2016}; using categorical
data~\cite{wang_high-resolution_2017}; or for \textit{cartoonization}
of photographs~\cite{Chen_2018_CVPR}.
    
Researchers have recently contributed to line drawings problem
pertaining to sketch simplification
\cite{simo-serra_learning_2016,simo-serra_mastering_2018}, line
drawing
colorization~\cite{furusawa_comicolorization_2017,kim_tag2pix_2019,zhang_two-stage_2018}
and line stylization~\cite{li_im2pencil_2019}.  Sketch
simplification~\cite{simo-serra_learning_2016,simo-serra_mastering_2018}
aims to clean up rough sketches by removing redundant lines and
connecting irregular lines. Researchers take a step ahead to develop a
tool~\cite{simo-serra_real-time_2018} to improve upon sketch
simplification by incorporating user input. It facilitates the users
to draw strokes indicating where they want to add or remove lines, and
then the model will output a simplified sketch.
\textit{Tag2Pix}~\cite{kim_tag2pix_2019} aims to use GANs based
architecture to colorize line drawing.  \textit{Im2pencil}
\cite{li_im2pencil_2019} introduce a two-stream deep learning model to
transform the line drawings to pencil drawings. Among the relevant
literature we studied, a recent work~\cite{zheng_learning_2020} seems
to be most related to our approach, where authors propose a method to
generate detailed and accurate artistic shadows from pairs of line
drawing sketches and lighting directions.

In general, it has been argued that models over the deep learning
paradigm handle more complex variations, but they also require large
training data.

\subsection{Sketch-based Modeling}
\label{sec:3d-projection}

\paragraph{Interactive Modeling}

\cite{Zeleznik:1996:SIS:237170.237238} introduced a set of rule-based
shortcuts with the aim of creating \textit{an environment for rapidly
  conceptualizing and editing approximate \textsc{3d}
  scenes}. Sketching was integrated soon into the system with
\textsc{teddy} and its relatives,
\cite{Igarashi:1999:TSI:311535.311602,Igarashi:2003:SMS:641480.641507,Mori:2007:PID:1276377.1276433},
with prime focus on inflated smooth shapes as first class citizens,
and sketch-based gestures for interactivity. Recently,
\textsc{smartcanvas} \cite{177} extended the works relying on planar
strokes.

\paragraph{Analytical Inference,}

for \textsc{3d} reconstruction is another popular approach to modeling
from sketches. With the context of vector line art, \cite{24791}
inferred simple curved shapes, and \cite{LIPSON1996651} showed
progress with complex compositions of planar surfaces. Recently, for
illuminating sketches,
\cite{shao:hal-00703202,Iarussi:2015:BRC:2774971.2710026} had shown
the effectiveness of normal-field inference through regularity cues,
while \cite{Xu:2015:ITS:2810210.2810212} used isophotes to infer the
normal-field.

\paragraph{Deep Learning Techniques,}

have more recently, shown substantial progress in inferring
\textsc{3d} models from sketches: a deep regression network was used
to infer the parameters of a bilinear morph over a face
mesh~\cite{Han:2017:DDL:3072959.3073629} training over the
\textit{FaceWarehouse} dataset \cite{6654137}; a generative
encoder-decoder framework was used to infer a fixed-size
point-cloud~\cite{Fan_2017_CVPR} from a real world photograph using a
distance metric, trained over \textit{ShapeNet} dataset
\cite{shapenet2015}; U-Net~\cite{ronneberger_u-net_2015}-like networks
were proven effective, to predict a volumetric model from a single and
multiple views~\cite{Delanoy:2018:SUM:3242771.3203197} using a diverse
dataset, but this dataset \textit{lacks pertinence to our task}.

\begin{figure*}[tbp]
  \centering
  \includegraphics[width=\linewidth]{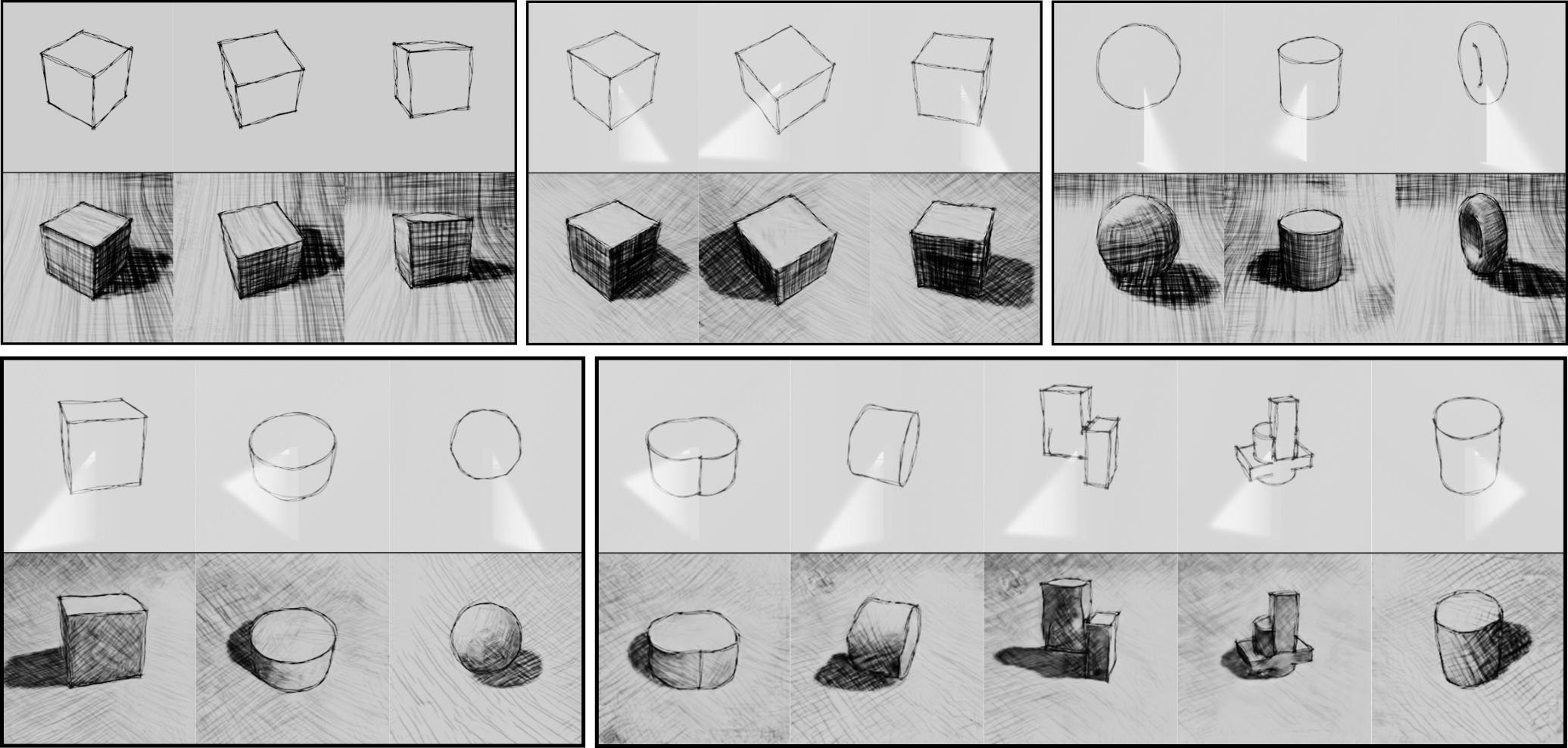}
  \caption[Progressive evaluations of \textsc{dm:wb}
  model]{Progressive evaluation of \textsc{dm:wb} (see
    \S~\ref{sec:experiment}) model varying camera pose, illumination,
    constituent geometry and texture. \emph{Clockwise from top left.}
    \textsc{pose}; \textsc{pose+lit}; \textsc{pose+lit+shap};
    \textsc{all}; \textsc{txr}. (Details at
    \S~\ref{sec:qualitative}.)}
  \label{fig:dm-wb-progressive}
\end{figure*}

\subsection{How we stand out}
\label{sub-sec:how_stand_out}

Deep learning applications has seeped into many research domains. The
profusion of data has helped the research community and has nicely
complemented deep learning algorithms. However, there does seem to be
lack of feasible, scalable, impressive in terms of both quality and
quantity, and user-interpretable dataset in the graphics community
specially artistic domain. There does seem to be a gap when we compare
these datasets with popular datasets suitable for deep learning
algorithms. For instance, ShadeSketch
dataset~\cite{zheng_learning_2020} \textit{doesn't offer much
  diversity} as it is limited to a specific domain and dataset has
comparitively less number of images.

In our approach, we aim to solve the problem of generating hatches for
sketches using our contextual image completion framework. The
framework ensures that the relevant context required in terms of
illumination information and textures is incorporated. We use
generative adversarial networks to ensure realistic generation of
hatch patterns that are indistinguishable from ground-truth hatch
patterns that are generated by using accurate ground-truth
\textsc{3d}. Note that in our inference procedure, we make use only of
the line drawings, the illumination and the texture context. We
\emph{do not} require access to \textsc{3d} or psuedo-\textsc{3d} in
order to generate the accurate hatch pattern. Further ways in which we
differ are as follows.

\textit{Firstly}, the problem of stylization inherently requires an
information-dense image to start with. For an artist, it is evidently
quicker as well as more intuitive, to create a basic line-art, than to
create a well rendered sketch; but line-art is sparse in nature.

\textit{Secondly}, the heuristic based methods although coarsely
capture the distribution, wherein different parameter values cater to
different class of problems, they are vulnerable in fringe cases. Deep
learning on the other hand has shown promise.

\textit{Finally,} in decision making, it is pivotal for the designers,
to control the stylization parameters of lighting and texture.

\section{Problem formulation}
\label{sec:methodology}


\begin{figure*}[htbp]
  \centering
  \includegraphics[width=\linewidth]{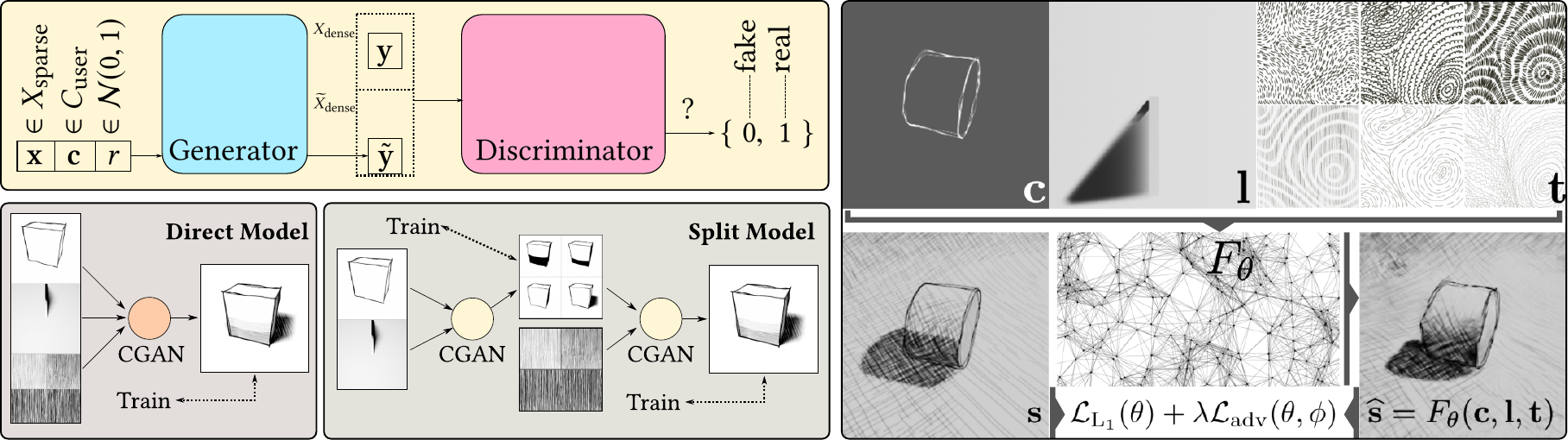}
  \caption{The overview of the our framework. \emph{Clockwise from
      top}. Illustration of structure of \textsc{gan} detailing
    injection of data from different modalities into the \textsc{cgan}
    framework; Further detailed view of direct model; Coarse sketch of
    a direct model,
    Eqs.~\eqref{eq:direct-recon-loss},\eqref{eq:direct-adv-loss},\eqref{eq:minimax};
    Analogous sketch of a split model,
    Eqs.~\eqref{eq:split-recon-loss},\eqref{eq:split-adv-loss},\eqref{eq:minimax}.}
  \label{fig:gan-framework}
\end{figure*}




  


The objective of this research is to learn a function
$\mathcal{F} : \textit{C} \times \textit{L} \times \textit{T}
\rightarrow \textit{S}$; where, \textit{C} represents the space of
hand-drawn contours for any \textsc{3d} geometry; \textit{L}
represents an illumination analogy shown for a canonical geometry;
\textit{T} is a set of textures representing different shades; and
\textit{S} is the space of completed sketches with shadows manifested
as hatch-patterns.

\subsection{Model}
\label{sec:model}

To this end, we leverage the \textsc{cgan}
\cite{NIPS2014_5423,DBLP:journals/corr/MirzaO14} framework. If
$F_\theta$ parameterised by $\theta$, be a family of functions
modelled as a deep network; $\mathbf{c} \in C$ be a line drawing,
$\mathbf{l} \in L$ be the illumination analogy, $\mathbf{t} \in T$ be
the set of tonal art maps (aka. textures), and $\mathbf{s} \in S$ be
the completed sketch, then the reconstruction loss is defined as in
Eq.~\eqref{eq:direct-recon-loss}.  For diversity in generation task,
$F_\theta$ is implemented with
dropouts~\cite{krizhevsky_imagenet_2012} as implicit noise. In order
to bring closer the model-generated sketches and real data
distribution, we define a discriminator $D_\phi$ parameterised by
$\phi$, as another family of functions modelled by a deep network. The
adversarial loss, is thus defined as in
Eq.~\eqref{eq:direct-adv-loss}. We hope the model to converge,
alternating the optimisation steps for the minimax in
Eq.~\eqref{eq:minimax}. We call this formulation as \textbf{direct
  model}, illustrated in Fig.~\ref{fig:gan-framework}.

In the same spirit, since data generation is cheap (see
\S~\ref{sec:dataset}), we see an opportunity to make use of the
illumination masks $\mathbf{m} \in M$ as an intermediate step for
supervision using a \textbf{split model}, for which the losses
$\mathcal{L}_\mathrm{L_1}$ and $\mathcal{L}_\mathrm{adv}$ in
Eq.~\eqref{eq:minimax} are formulated as in
Eqs.~\eqref{eq:split-recon-loss}~and~\eqref{eq:split-adv-loss}.

\begin{align}
  \theta^* &= \begin{aligned}[t]\label{eq:minimax}
    &\arg \min_\theta \max_\phi \mathbb{E}_{
      \mathbf{c},\mathbf{l},\mathbf{t},\mathbf{s}\sim\mathrm{data}
    }\\
    &\left[ \mathcal{L}_\mathrm{L_1}(\theta) + \lambda
      \mathcal{L}_\mathrm{adv}(\theta, \phi) \right]
  \end{aligned}\\
  \textrm{\bfseries Direct Model:} \notag \\  
  \widehat{\mathbf{s}} &= F_\theta(\mathbf{c}, \mathbf{l},
                         \mathbf{t}) \notag\\
  \mathcal{L}_\mathrm{L_1}(\theta) &= \left\lVert \mathbf{s} - \widehat{\mathbf{s}}
                                     \right\rVert_1 \label{eq:direct-recon-loss} \\
  \mathcal{L}_\mathrm{adv}(\theta, \phi)
                       &= \log(D_\phi(\mathbf{s})) +
                         \log(1-D_\phi(\widehat{\mathbf{s}}))  \label{eq:direct-adv-loss}  \\
  \textrm{\bfseries Split Model:} \notag \\  
  \widehat{\mathbf{m}} &= F_\theta^{(1)}(\mathbf{c}, \mathbf{l})
                         \notag\\
  \widehat{\mathbf{s}} &= F_\theta^{(2)}(\widehat{\mathbf{m}}, \mathbf{t})
                         \notag\\
  \mathcal{L}_\mathrm{L_1}(\theta) &= \left\lVert \mathbf{m} -
                              \widehat{\mathbf{m}} \right\rVert_1 +
                              \left\lVert \mathbf{s} -
                              \widehat{\mathbf{s}}
                              \right\rVert_1 \label{eq:split-recon-loss}\\ 
  \mathcal{L}_\mathrm{adv}^{(1)}(\theta, \phi) &= \log(D_\phi^{(1)}(\mathbf{m})) +
                                          \log(1-D_\phi^{(1)}(\widehat{\mathbf{m}}))
                                          \notag\\
  \mathcal{L}_\mathrm{adv}^{(2)}(\theta, \phi) &= \log(D_\phi^{(2)}(\mathbf{s})) +
                                          \log(1-D_\phi^{(2)}(\widehat{\mathbf{s}}))
                                          \notag\\
  \mathcal{L}_\mathrm{adv}(\theta, \phi) &= \mathcal{L}_\mathrm{adv}^{(1)}(\theta,
                                    \phi) +
                                    \mathcal{L}_\mathrm{adv}^{(2)}(\theta,
                                    \phi) \label{eq:split-adv-loss}
\end{align}


\subsection{Architecture}
\label{sec:architecture}

Akin to earlier attempts~\cite{isola_image--image_2016}, we utilise
U-Net~\cite{ronneberger_u-net_2015} with $\approx 12\mathrm{M}$ as our
base architecture. The only main difference being the use of
\emph{multi-modal bounding conditions} that are infused into the
network by varying the number of input channels, namely the
\textit{sparse object outline sketch}, \textit{illumination hint}, and
\textit{texture} (See Fig.~\ref{fig:gan-framework}). We experiment
with alternate formulations of the model so that the complete sketch
is predicted using a direct model
Eqs.~\eqref{eq:direct-recon-loss},\eqref{eq:direct-adv-loss}, or by
predicting an extra intermediate representation of illumination masks
using a split model
Eq.~\eqref{eq:split-recon-loss}~\eqref{eq:split-adv-loss}. The
discriminator is designed as a PatchGAN based classifier, which
determines whether $N \times N$ patch in an image is real or fake.

Inspired by recent success of self-attention in adversarial
nets~\cite{zheng_learning_2020}, we also use a
squeeze-and-excitation~\cite{hu_squeeze-and-excitation_2018} based
architecture for our model, $F_\theta$ in
Eqs.~\eqref{eq:direct-recon-loss},\eqref{eq:direct-adv-loss}.

\subsection{Remarks}
\label{sec:problem-remarks}

We propose a conditional \textsc{gan} framework to solve for the
task. We adopt the baseline architecture
\cite{isola_image--image_2016}, and further implement a model inspired
by a recent advancement as an ablation study. However, our
architecture is different and stands out from prior works in the way
how muti-modal bounding conditions are being infused into the
network. The model so designed can be trained in an end-to-end manner
while satisfying the desired constraints.

As a general notion, deep neural networks are difficult to train and
are sensitive to the choice of hyper-parameters. 
But our model doesn't introduce additional complex components, and
thus allows us to take advantage of knowing the hyper-parameters of
underlying base architecture. This also results in a conceptually
clean and expressive model that relies on principled and
well-justified training procedures. This is, to the best of our
knowledge, the first attempt to solve the problem of generating
hatches for sketches while incorporating multiple modalities as
constraints.

\section{Dataset}
\label{sec:dataset}


\begin{figure}[tb]
  \centering

  \label{fig:tams}
  \includegraphics[width=\columnwidth]{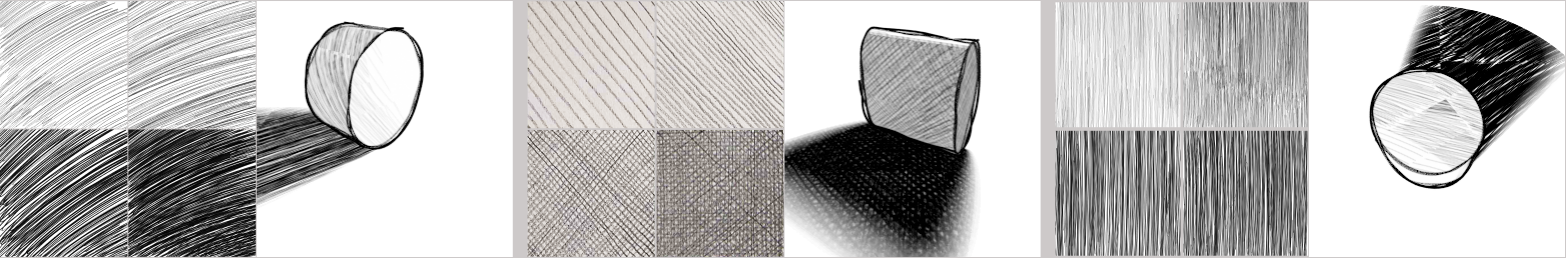}
  \caption{Three sketches rendered using three different objects with
    \textsc{tam}'s implemented using \textsc{blender}.}

  \vspace{0.15in}
  \label{fig:dataset-single-point}
  \def\svgwidth{\linewidth}
  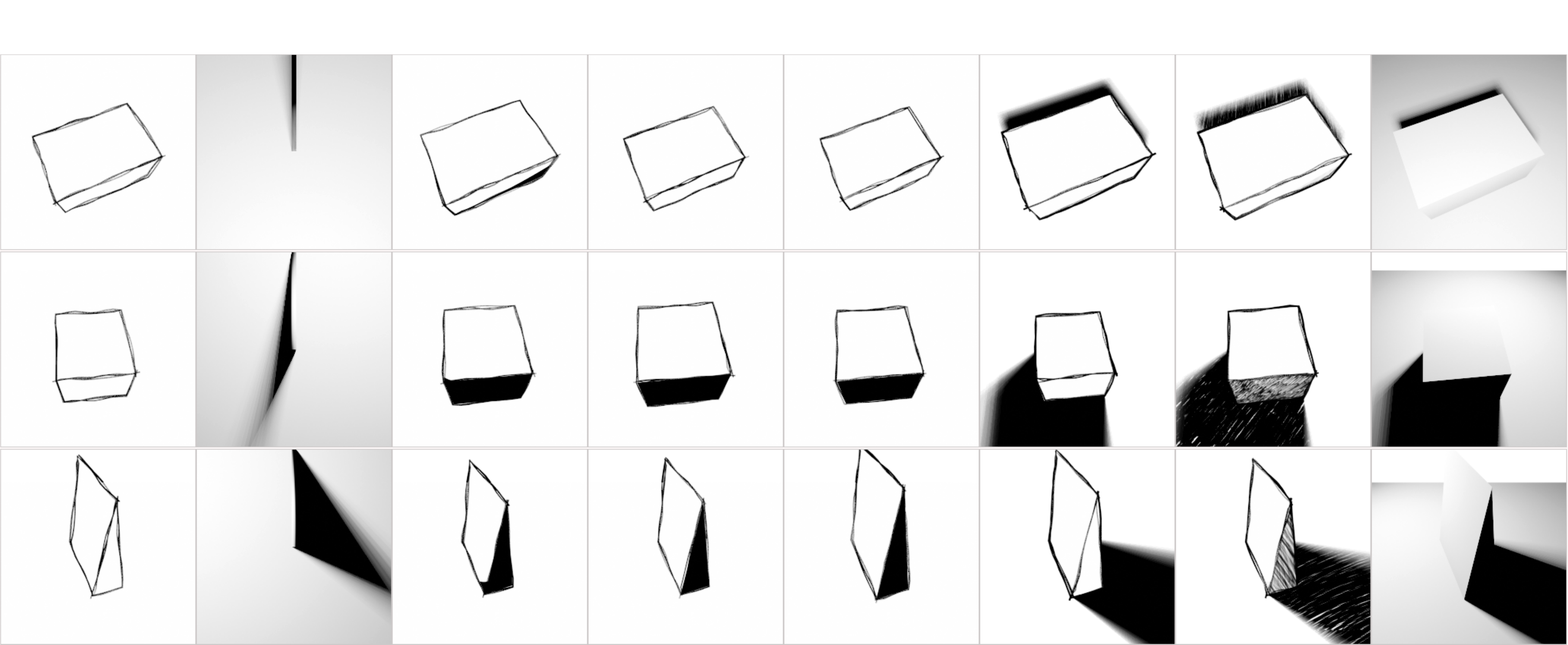
  \caption[Examples from the dataset.]{Few examples from the
    dataset. \emph{Columns left to right}. \textsc{cnt}: Contour;
    \textsc{ill}: Illumination; \textsc{hi}: Highlight mask;
    \textsc{mid}: Midtone mask; \textsc{sha}: Shade mask (on object);
    \textsc{shw}: Shadow mask; \textsc{sk}: Sketch render;
    \textsc{dif}: Diffuse render.}

  \vspace{0.15in}
  \label{fig:sundial}
  \includegraphics[width=\columnwidth]{images/sundial-inspiration.png}
  \caption{Gnomon in a sundial is used as canonical object to capture
    illumination information. \textit{Left to right:} A
    sundial~(\textit{Courtesy:} \textit{liz west}
    {\small\url{https://flic.kr/p/EWBd4}}); Perspective render of a
    gnomon; A top view of the same.}

\end{figure}



We introduce the \textsc{shad3s} dataset (
Fig.~\ref{fig:dataset-single-point}), where each data-point contains:
\textit{a)} a contour drawing of a \textsc{3d} object; \textit{b)} an
illumination analogy; \textit{c)} 3 illumination masks over
\textit{a}, namely \textit{highlights, midtones} and \textit{shades};
\textit{d)} a shadow mask on the ground; \textit{e)} a diffuse render;
and \textit{f)} a sketch render. Additionally, it contains a catalogue
of $6$ textures.


\paragraph{Geometry and render}

This is a fully synthetic dataset created from scratch
using the following steps: \emph{a)} Create a
Constructive Solid Geometry (\textsc{csg})-graph using
simple grammar over a finite set of eulidean solids;
\emph{b)} Create a mesh from \textsc{csg}; \emph{c)}
Render the mesh. The illumination masks were created by
thresholding a diffuse
shader. Blender's~\cite{blender_online_community_blender_2019}
official freestyle
tool~\cite{the_blender_foundation_freestyle_2019} was
used to render the contours. And sketch renders were
created after implementing the tonal art
maps~\cite{praun_real-time_2001} (see
Fig.~\ref{fig:tams}).

Since the geometry is also procedurally generated on the fly, we take
the opportunity to progressively increase the complexity of geometry
by varying the upper-bound of number of solids in the composition
between 1 and 6. This assists in the progressive evaluation of the
models.

\paragraph{Illumination and background}

Additionally, we also froze the information of illumination conditions
in the form of a diffuse render of a canonical object resembling a
gnomon (of a sundial). The rationale behind using this style of
information is to provide information to image domain. We also
attempted to use a background filled with hatch patterns, instead of
transparent background, and analysed its effects on the model
performance detailed out in \S~\ref{sec:experiment}.

\paragraph{Textures as Tonal art maps}

For the purpose of this research, we follow the principle of tonal art
maps \cite{praun_real-time_2001}, albeit in a low-fidelity version. To
approximate the effect, we create an 6-tone cel-shading \cite{942087},
including the pure white and pure black. For the rest of the four
shades we map one texture corresponding to each tone on the
object.

The similarity to tonal art maps lies in the fact that any hatch
pattern map for a given tone should also be visible in all the tones
darker than itself.  To this effect, we collect samples of high
resolution tonal art maps apriori, from artists. A random crop of
these images is used by the model as input conditions (see
Fig.~\ref{fig:tams})



\paragraph{All in all}

we create \textit{six subsets}, each restricting the maximum number of
solids in a \textsc{csg} composition of a scene to be from \textit{1
  through 6}. Each subset is created from $\sim 1024$ distinct
scenes. Each scene is rendered from $\sim 64$ different camera
poses. The dataset thus contains $\sim 2^{16}\times 6$ data points
with a resolution of $\sim 256\times 256$.

\section{Experimentation and results}
\label{sec:experiment}

\begin{figure}[htbp]
  \centering
  \def\svgwidth{\linewidth}
  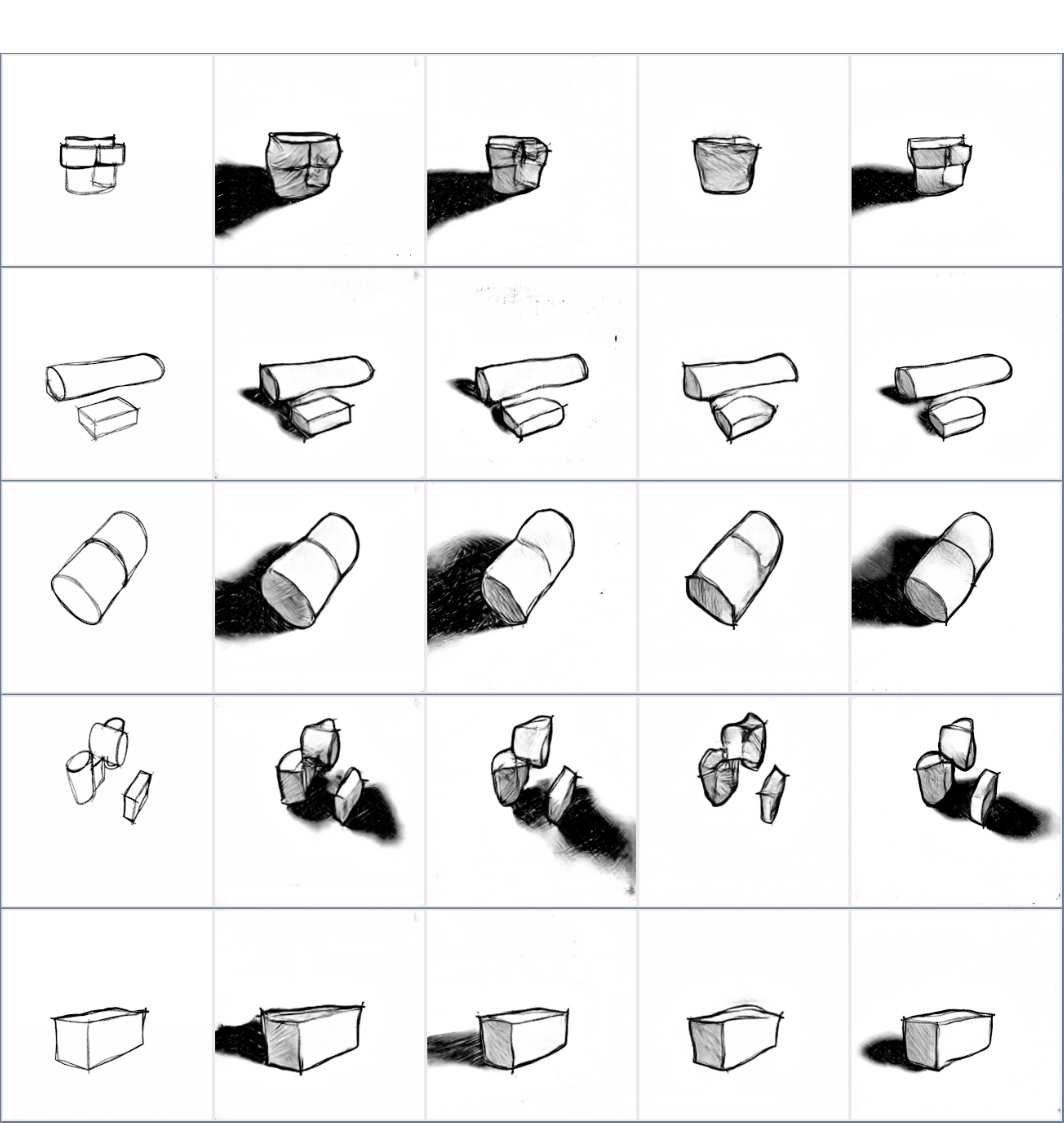
  \caption[Comparative results for all models]{Comparative results for
    all models. \emph{Columns left to right}. Input contour drawing;
    Corresponding evaluation with \textsc{dm}; With \textsc{sp}; With
    \textsc{sp:ws}; With \textsc{se}.}
  \label{fig:eval-all}
\end{figure}

\begin{figure}[htbp]
  \centering
  \def\svgwidth{\linewidth}
  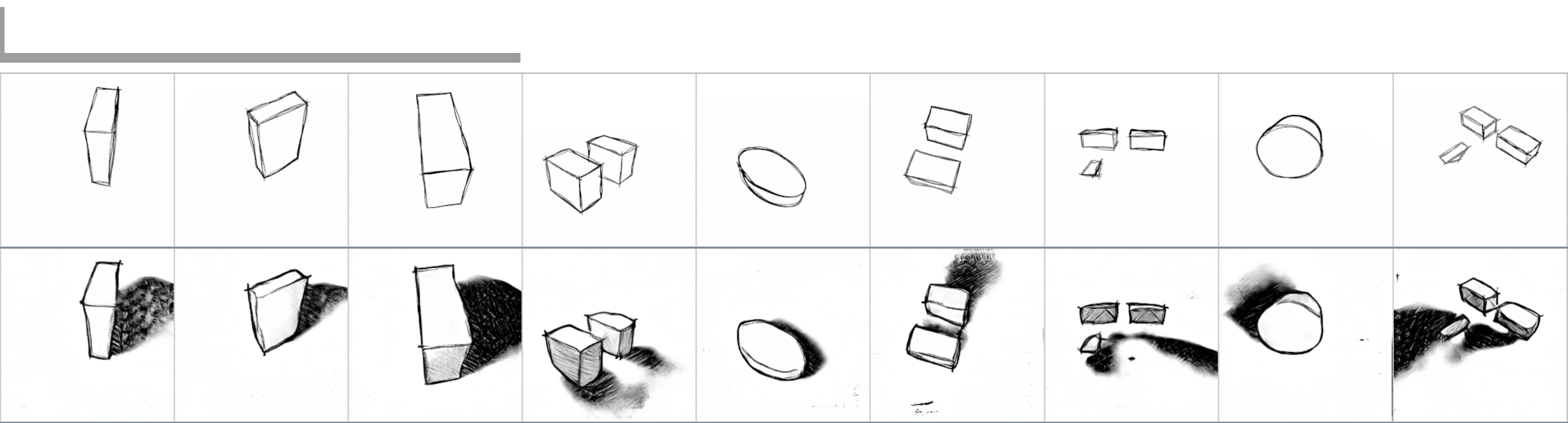
  \caption[Progressive results for \textsc{dm}]{Progressive results
    for \textsc{dm}. Left three with single solid on scene; Middle
    three with upto two solids; and Rightmost three with upto three
    solids.}
  \label{fig:dm-progressive}
\end{figure}

\begin{table}
  \caption{Quantitative evaluation using \textsc{psnr},
    \textsc{ssim}, Inference time (in ms), followed by
    inception scores of model predictions, juxtaposed
    against those of the dataset.}
\label{tab:quantitative}
\centering
\begin{tabular}{lrrrrr}
  \toprule
  \textsc{model} & \textsc{psnr}  & \textsc{ssim} & Time & Pred. \textsc{is} & \textsc{gt is}  \\ \midrule
  \textsc{dm}  & 55.72 & 0.347 & 13 & 4.25 & 5.51 \\
  \textsc{sp}  &  55.48 & 0.349 & 33 & 4.12 & 5.51\\
  \textsc{se} & 55.90 & 0.376 & 426 & 4.35 & 5.51 \\
  \textsc{sp:ws}  & 58.92 & 0.287 & 36 & 3.71 & 5.27\\
  \bottomrule
\end{tabular}

\end{table}

\begin{table}
  \caption{Progressive improvement in inception scores
    against an increase in dataset complexity}
\label{tab:dataset}
\centering
\begin{tabular}{l|rrrrrr}
  \toprule
  Max objects & $1$ & $2$ & $3$ & $4$ & $5$ & $6$  \\ \midrule
  \textsc{gt is} & $3.36$ & $3.94$ & $4.13$ & $4.86$ & $5.15$ & $5.51$ \\
  
  \bottomrule
\end{tabular}
\end{table}

To test our hypothesis we performed experiments on the three models,
namely \emph{a)} the direct model over U-Net architecture
(\textsc{dm}); \emph{b)} the split model over U-Net architecture
(\textsc{sp}); \emph{c)} the split model over squeeze-and-excitation
architecture (\textsc{se}).  As an extension, we also studied the
direct model trained over a dataset with background (\textsc{dm:wb}),
and the split model trained over a dataset without shadows on ground
(\textsc{sp:ws}).  Further, the results are analysed qualitatively
(\S~\ref{sec:qualitative}), quantitatively (\S~\ref{sec:quantitative})
and for generalizability of the model (\S~\ref{sec:generalizability}).


\subsection{Qualitative Evaluation}
\label{sec:qualitative}

Initially the a \textsc{dm} model was trained with low resolution
datasets, starting with single solid in a scene, through to a
combination of upto three solids in a scene. The indicative results in
Fig.~\ref{fig:dm-progressive},show that with progressive increase in
the complexity of scene, the model struggles to keep up with the
expected response.

To illustrate the strength of our experiments, we present the
qualitative results of the following progressive analysis of our
\textsc{dm:wb} model with increasingly complex scenes, as shown in
Fig.~\ref{fig:dm-wb-progressive}. The model visibly learns to shade
well for given lighting conditions, and standard primitives over a
canonical texture. The figure is coded as follows,

\noindent{}\textsc{pose}: \emph{Pose variations} of a
cube, with a canonical scene.

\noindent{}\textsc{pose+lit}: \emph{Variations in
  lighting} and pose of a cube, combined with a
canonical texture.

\noindent{}\textsc{pose+lit+shap}: \emph{Variations in
  lighting}, pose and shape (single primitive solid in scene), with a
canonical texture.

\noindent{}\textsc{txr}: \emph{Variations in lighting},
pose, shape (single primitive solid in scene) and texture.

\noindent{}\textsc{all}: \emph{Variations in lighting},
pose, texture and complex compositions (upto six solids combined in a
scene.)

Further qualitative comparison of all the five models, are presented
in Fig.~\ref{fig:eval-all}. The nuances of shading with complex
solids, and the consequent mistakes, that the model incurs, may escape
an untrained eye, solely because, the hatching and shading is
acceptable at a coarse level. The models trained without background
also perform well but qualitatively seem far from the performance of a
model trained with background. Self occlusion and self shadows pose a
major challenge.


\subsection{Quantitative analysis}
\label{sec:quantitative}

Inception score has been a widely used evaluation metric for
generative models. This metric was shown to correlate well with human
scoring of the realism of generated
images~\cite{salimans2016improved}. We see in Table~\ref{tab:dataset}
that as expected the metric increases with increase in scene
complexity, indicating progressively more diversity through the
experiment.

We evaluate the models, \textsc{dm, sp, sp:ws, se}, and results are
summarised under Table~\ref{tab:quantitative}. In our case we use the
\textsc{npr} rendered images as a reference to the generated sketch
completions. The inception scores of the images generated by our
models, 
are comfortably close to the scores for ground truth
(\textsc{gt}). Furthermore, the metrics \textsc{psnr, ssim} and
\textsc{is} exhibit mild perturbations amongst the models trained with
shadow, whereas a sharp change in case of the model trained without
shadows.  We see a steep decline in the diversity (\textsc{ssim} and
\textsc{is}), and a sharp ascent in \textsc{psnr}, both of which can
be explained by the absence of shadows, that on one hand limit the
scope of expressivity and \textit{diversity}; and on the other, limit
the scope of discrepancy from \textsc{gt} and hence \textit{noise},
improving the count of positive signals.


\subsection{Generalizability}
\label{sec:generalizability}

\begin{figure}[htbp]
  \centering
  \def\svgwidth{\linewidth}
  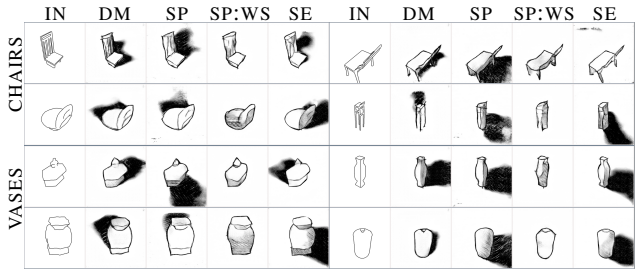
  \caption[Generalizability results]{Generalizability results ---
    chairs in top two rows; and vases in the next. \emph{Set of five
      columns left to right.} Input contours; Evaluation with
    \textsc{dm}; With \textsc{sp}; With \textsc{sp:ws}; With
    \textsc{se}.}
  \label{fig:genz}
\end{figure}

We deploy two methods to verify a model's tendency to overfit, and/or
its ability to generalize. One is to run on other publicly available
datasets. To this end, we utilize the class of objects in chairs and
vases from the ShapeNet~\cite{shapenet2015}. And the other method we
deploy is by user evaluation (see~\S~\ref{sec:user-evaluation}). The
representative results of qualitative evaulation can be seen in
Fig.~\ref{fig:genz}.

\subsection{Limitations}
\label{sec:limitations}

Although the model is able to predict even for human sketches, unseen
as well as totally different from the seen dataset, they are barely
satisfactory. We have seen here that the model at times modifies the
contours, which raises a question of fidelity in production
mode. Also, to the best of our knowledge, there seems to be a lack of
popularly accepted metric for measuring the levels of realism and
diversity in the human sketches. Hence, we borrow the metrics of
Inception score\cite{salimans2016improved}, \textsc{psnr, ssim} from
optical domain as a proxy to examine and verify a similar trend for
our dataset.





\section{User Evaluation}
\label{sec:user-evaluation}

\begin{figure}[htbp]
  \centering
  \def\svgwidth{\linewidth}
  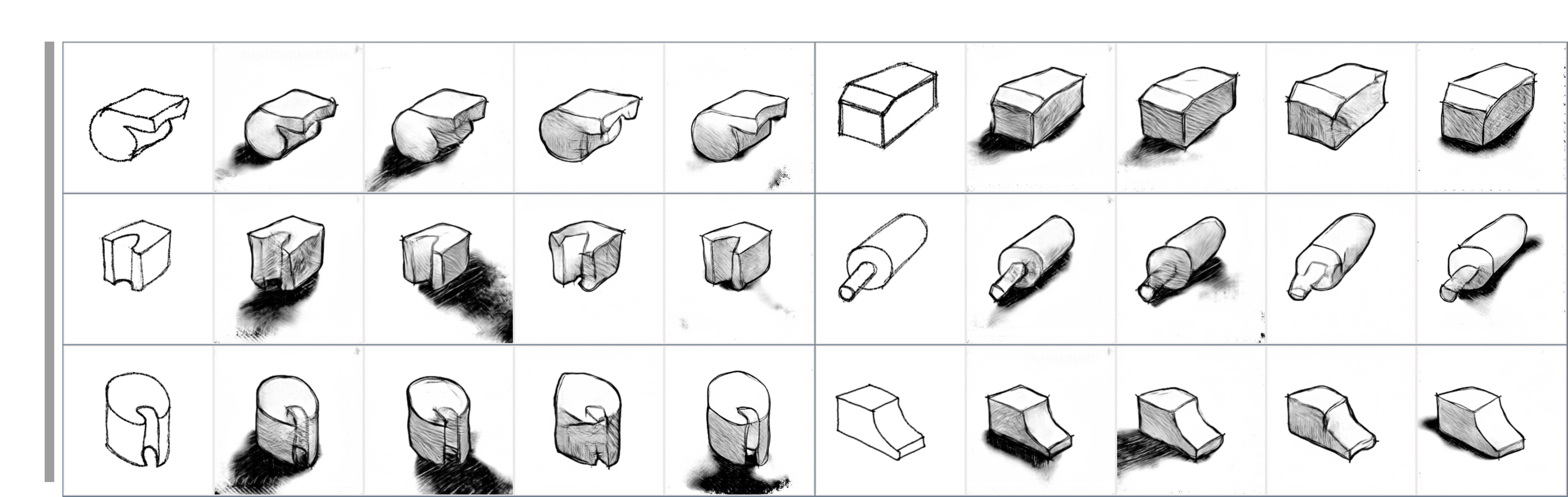
  \def\svgwidth{\linewidth}
  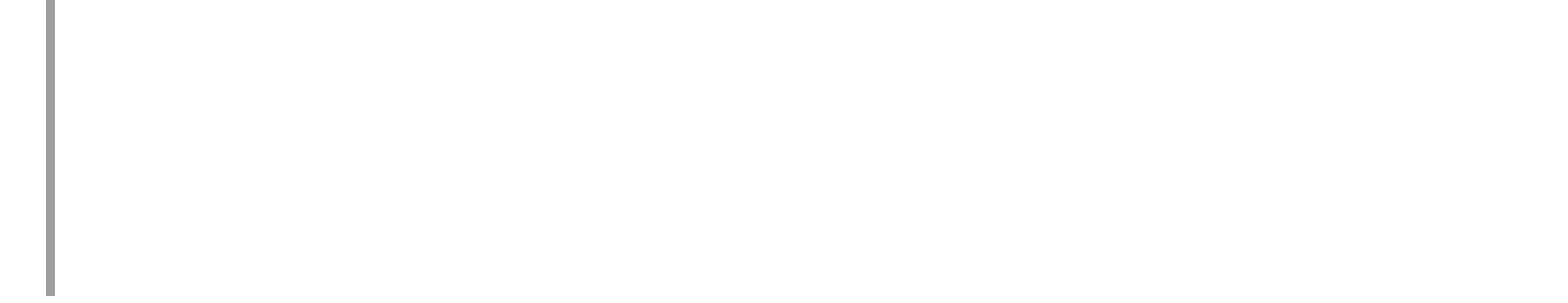
  \def\svgwidth{\linewidth}
  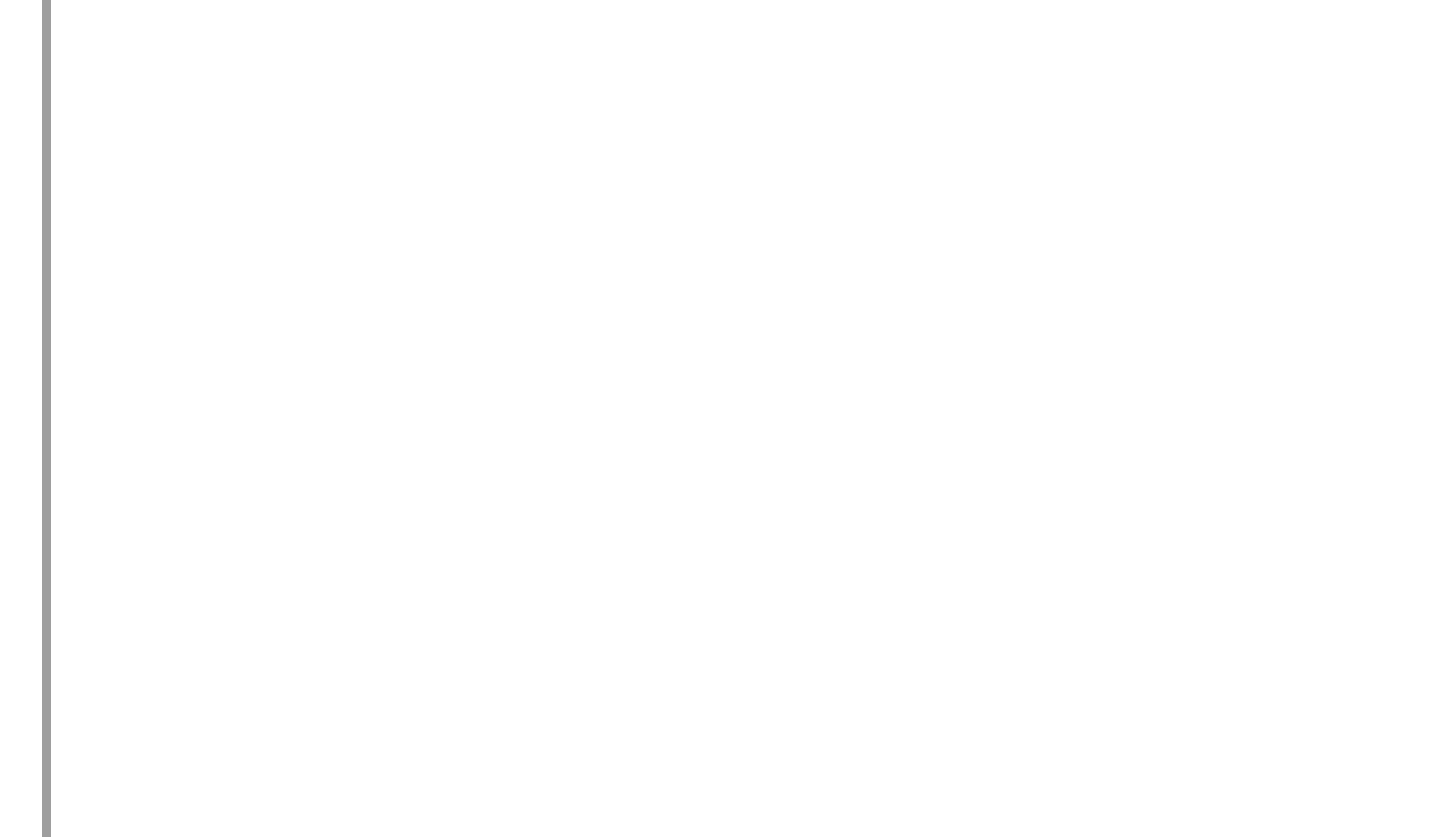
  \caption[User evaluation results]{User evaluation
    results. \emph{Blocks top to bottom}. Geometric shapes; Similar
    shapes with geometric and organic variations; Organic
    shapes. \emph{Set of five columns left to right.} Input contours;
    Evaluation with \textsc{dm}; With \textsc{sp}; With
    \textsc{sp:ws}; With \textsc{se}.}
  \label{fig:user-eval}
\end{figure}

\begin{figure}[htbp]
  \centering
  \def\svgwidth{\linewidth}
  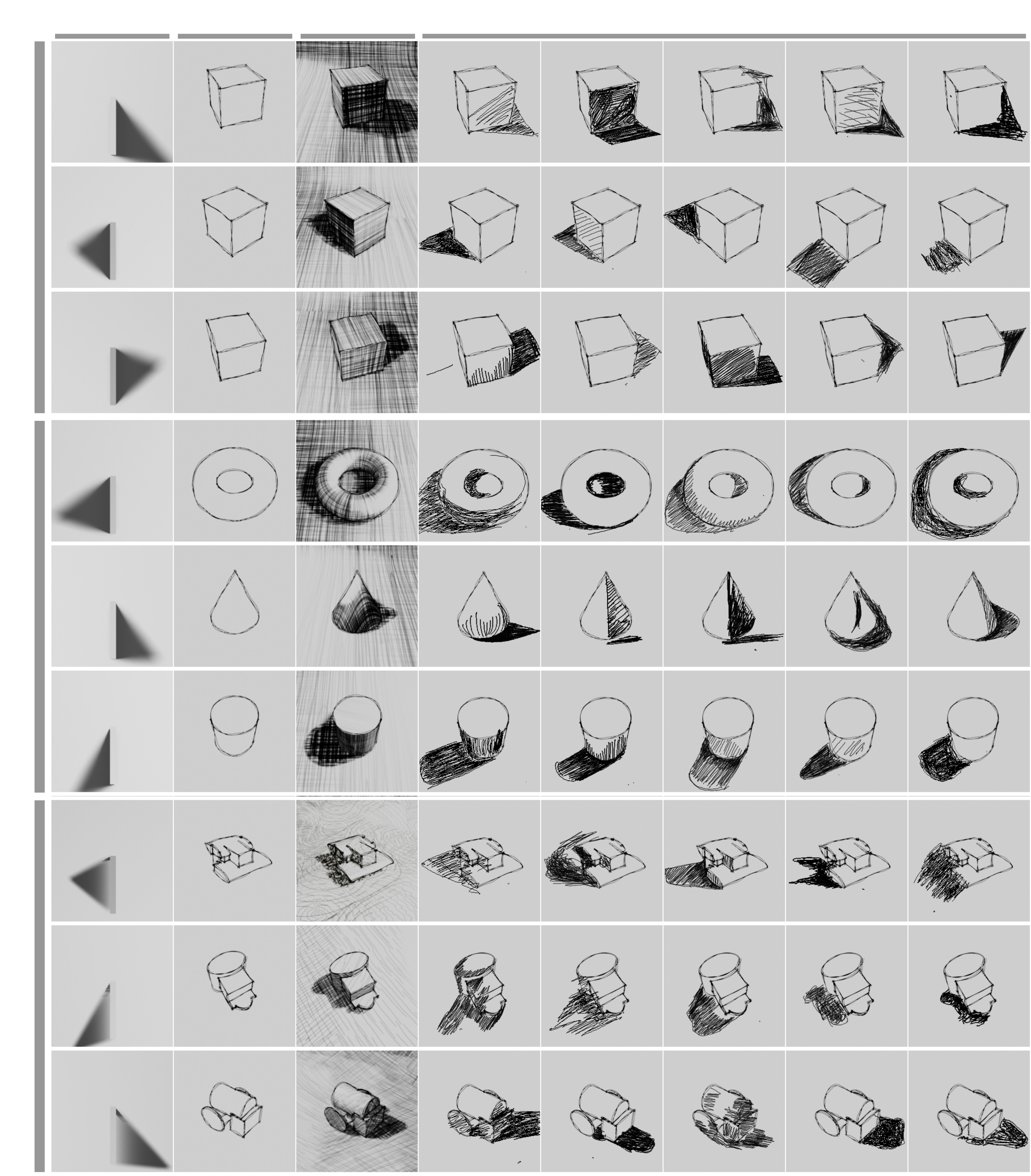
  \caption[User task completion]{Selected responses from user task
    completion (see \S~\ref{sec:user-evaluation}). \emph{Blocks top to
      bottom}. With only \textsc{cubes} in scene; With
    \textsc{primitives} only; With \textsc{complex}
    geometries. \emph{Columns left to right.}  \textsc{ill}:
    Illumination hint; \textsc{cnt}: Contours; \textsc{sk}:
    \textsc{gt} sketch for reviewers reference; Next five columns show
    users responses.}
  \label{fig:user-hand}
\end{figure}

We prepared a tool called \textsc{dhi}---deeply hatch it, for a user
to evaluate the model in a real world scenario. The tool leverages the
widely popular \textsc{gimp} program over a pen-touch display to
provide the \emph{real-world} digital drawing experience. There is
another \textsc{tk}-based applet which allows for choosing the
illumination and texture. For consistency, the illumination images
were computed in real-time on client-side using the blender backend.
The image from the user-end \emph{(client)} is sent to the compute
server for inference triggered by a click on the result window on the
\textsc{tk} applet, where the response is displayed.

We have classified the interactions as \emph{a)} casual, and \emph{b)}
involved. The former was more of a rendezvous with the tool and
model. In order to get familiar with the tool, the user was guided by
a mentor for a couple of drawings, and later left alone to engage with
the tool, hoping for her to develop familiarity and find her comfort
zone .

The time of participant engagement varied between 5-43 minutes
\((\mu=21.47,\, \sigma=5.24)\). The sample size was 55 users with
their age varying between 12-57 years \((\mu=26.15,\,
\sigma=7.04)\). The results of this evaluation are shown in
Fig.~\ref{fig:user-eval}. More impressive results from model trained
with background \textsc{dm:wb} have been shown in
Fig.~\ref{fig:teaser}.

We also conducted a more involved group based task completion exercise
with 19 volunteers having art/design background, split into groups
of 4. This exercise was conducted in three phases and lasted a couple
of hours either side of the lunch. 

The first phase was an individual level task. A user was required to
complete a sketch instead of the computer in a threshold of 40 sec;
the threshold had been chosen by consensus. A total of 21 samples were
presented, split into three levels of difficulty: \emph{a)} the first
being the nascent level where only a cube is looked at from different
poses; \emph{b)} the second consisting a simple geometry like
euclidean solids and tori; and \emph{c)} the last were combination of
multiple solids generated procedurally. The results shown in
Fig.~\ref{fig:user-hand} indicate that the users seem to have
understood the problem principally. The second phase was akin to the
casual interaction as described earlier. The last phase, was a group
task where a group was required to utilize the tool to create a stop
motion picture, one frame at a time.

We encourage the reader to view the supplementary
video\footnote{Supplementary video is accessible on project page,
  \ppg} for the results of this exercise, evaluated against different
models.

\section{Conclusion}
\label{sec:conclusion}

We have shown the use of \textsc{cgan} in our pipeline for conditional
generation of hatch patterns corresponding to line-art representing
\textsc{3d} shapes, by synthesising a large training set, and
have portrayed its generalizability over a wide variety of real world
user input.

We started with a basic model called \textsc{dm} and experimented with
architectures and models with varying complexity, assuming that model
capacity was limiting to get more satisfactory results. However, that
wasn't the case as suggested by our investigations and as seen from
the results; there is no clear winner, in general. From a utilitarian
perspective, we should hence suggest the use of simplest and most
efficient model, ie. \textsc{dm}.

This is nascent research area, which opens up potential investigation
into higher fidelity hatching patterns. Recent advances from the deep
learning community have the potential to be explored to improve the
results.

{\small
\bibliographystyle{ieee_fullname}
\bibliography{egbib}
}

\end{document}

%% file: images/dataset-summary-pdf.pdf_tex
\begingroup%
  \makeatletter%
  \providecommand\color[2][]{%
    \errmessage{(Inkscape) Color is used for the text in Inkscape, but the package 'color.sty' is not loaded}%
    \renewcommand\color[2][]{}%
  }%
  \providecommand\transparent[1]{%
    \errmessage{(Inkscape) Transparency is used (non-zero) for the text in Inkscape, but the package 'transparent.sty' is not loaded}%
    \renewcommand\transparent[1]{}%
  }%
  \providecommand\rotatebox[2]{#2}%
  \newcommand*\fsize{\dimexpr\f@size pt\relax}%
  \newcommand*\lineheight[1]{\fontsize{\fsize}{#1\fsize}\selectfont}%
  \ifx\svgwidth\undefined%
    \setlength{\unitlength}{2066.00028992bp}%
    \ifx\svgscale\undefined%
      \relax%
    \else%
      \setlength{\unitlength}{\unitlength * \real{\svgscale}}%
    \fi%
  \else%
    \setlength{\unitlength}{\svgwidth}%
  \fi%
  \global\let\svgwidth\undefined%
  \global\let\svgscale\undefined%
  \makeatother%
  \begin{picture}(1,0.41142297)%
    \lineheight{1}%
    \setlength\tabcolsep{0pt}%
    \put(0,0){\includegraphics[width=\unitlength,page=1]{images/dataset-summary-pdf.pdf}}%
    \put(0.03350152,0.38759484){\makebox(0,0)[lt]{\lineheight{1.25}\smash{\begin{tabular}[t]{l}\small\textsc{cnt}\end{tabular}}}}%
    \put(0.17016576,0.38759484){\makebox(0,0)[lt]{\lineheight{1.25}\smash{\begin{tabular}[t]{l}\small\textsc{ill}\end{tabular}}}}%
    \put(0.29674661,0.38759484){\makebox(0,0)[lt]{\lineheight{1.25}\smash{\begin{tabular}[t]{l}\small\textsc{hi}\end{tabular}}}}%
    \put(0.40401437,0.38759484){\makebox(0,0)[lt]{\lineheight{1.25}\smash{\begin{tabular}[t]{l}\small\textsc{mid}\end{tabular}}}}%
    \put(0.53518253,0.38759484){\makebox(0,0)[lt]{\lineheight{1.25}\smash{\begin{tabular}[t]{l}\small\textsc{sha}\end{tabular}}}}%
    \put(0.65503022,0.38759484){\makebox(0,0)[lt]{\lineheight{1.25}\smash{\begin{tabular}[t]{l}\small\textsc{shw}\end{tabular}}}}%
    \put(0.79593708,0.38759484){\makebox(0,0)[lt]{\lineheight{1.25}\smash{\begin{tabular}[t]{l}\small\textsc{sk}\end{tabular}}}}%
    \put(0.91293752,0.38759484){\makebox(0,0)[lt]{\lineheight{1.25}\smash{\begin{tabular}[t]{l}\small\textsc{dif}\end{tabular}}}}%
  \end{picture}%
\endgroup%

%% file: images/eval-all-pdf.pdf_tex
\begingroup%
  \makeatletter%
  \providecommand\color[2][]{%
    \errmessage{(Inkscape) Color is used for the text in Inkscape, but the package 'color.sty' is not loaded}%
    \renewcommand\color[2][]{}%
  }%
  \providecommand\transparent[1]{%
    \errmessage{(Inkscape) Transparency is used (non-zero) for the text in Inkscape, but the package 'transparent.sty' is not loaded}%
    \renewcommand\transparent[1]{}%
  }%
  \providecommand\rotatebox[2]{#2}%
  \newcommand*\fsize{\dimexpr\f@size pt\relax}%
  \newcommand*\lineheight[1]{\fontsize{\fsize}{#1\fsize}\selectfont}%
  \ifx\svgwidth\undefined%
    \setlength{\unitlength}{1303.99996948bp}%
    \ifx\svgscale\undefined%
      \relax%
    \else%
      \setlength{\unitlength}{\unitlength * \real{\svgscale}}%
    \fi%
  \else%
    \setlength{\unitlength}{\svgwidth}%
  \fi%
  \global\let\svgwidth\undefined%
  \global\let\svgscale\undefined%
  \makeatother%
  \begin{picture}(1,1.05567853)%
    \lineheight{1}%
    \setlength\tabcolsep{0pt}%
    \put(0,0){\includegraphics[width=\unitlength,page=1]{images/eval-all-pdf.pdf}}%
    \put(0.09040881,1.02000879){\makebox(0,0)[t]{\lineheight{1.25}\smash{\begin{tabular}[t]{c}\textsc{in}\end{tabular}}}}%
    \put(0.29272877,1.02000879){\makebox(0,0)[t]{\lineheight{1.25}\smash{\begin{tabular}[t]{c}\textsc{dm}\end{tabular}}}}%
    \put(0.49504869,1.02000879){\makebox(0,0)[t]{\lineheight{1.25}\smash{\begin{tabular}[t]{c}\textsc{sp}\end{tabular}}}}%
    \put(0.69736859,1.02000879){\makebox(0,0)[t]{\lineheight{1.25}\smash{\begin{tabular}[t]{c}\textsc{sp:ws}\end{tabular}}}}%
    \put(0.89968854,1.02000879){\makebox(0,0)[t]{\lineheight{1.25}\smash{\begin{tabular}[t]{c}\textsc{se}\end{tabular}}}}%
  \end{picture}%
\endgroup%

%% file: images/progressive-pdf.pdf_tex
\begingroup%
  \makeatletter%
  \providecommand\color[2][]{%
    \errmessage{(Inkscape) Color is used for the text in Inkscape, but the package 'color.sty' is not loaded}%
    \renewcommand\color[2][]{}%
  }%
  \providecommand\transparent[1]{%
    \errmessage{(Inkscape) Transparency is used (non-zero) for the text in Inkscape, but the package 'transparent.sty' is not loaded}%
    \renewcommand\transparent[1]{}%
  }%
  \providecommand\rotatebox[2]{#2}%
  \newcommand*\fsize{\dimexpr\f@size pt\relax}%
  \newcommand*\lineheight[1]{\fontsize{\fsize}{#1\fsize}\selectfont}%
  \ifx\svgwidth\undefined%
    \setlength{\unitlength}{2326.00012207bp}%
    \ifx\svgscale\undefined%
      \relax%
    \else%
      \setlength{\unitlength}{\unitlength * \real{\svgscale}}%
    \fi%
  \else%
    \setlength{\unitlength}{\svgwidth}%
  \fi%
  \global\let\svgwidth\undefined%
  \global\let\svgscale\undefined%
  \makeatother%
  \begin{picture}(1,0.26999139)%
    \lineheight{1}%
    \setlength\tabcolsep{0pt}%
    \put(0,0){\includegraphics[width=\unitlength,page=1]{images/progressive-pdf.pdf}}%
    \put(0.01075073,0.24562428){\makebox(0,0)[lt]{\lineheight{1.25}\smash{\begin{tabular}[t]{l}\textsc{single solid}\end{tabular}}}}%
    \put(0.34482713,0.24687923){\makebox(0,0)[lt]{\lineheight{1.25}\smash{\begin{tabular}[t]{l}\textsc{upto two}\end{tabular}}}}%
    \put(0.67886222,0.24687923){\makebox(0,0)[lt]{\lineheight{1.25}\smash{\begin{tabular}[t]{l}\textsc{upto three}\end{tabular}}}}%
    \put(0,0){\includegraphics[width=\unitlength,page=2]{images/progressive-pdf.pdf}}%
  \end{picture}%
\endgroup%

%% file: images/genz-pdf.pdf_tex
\begingroup%
  \makeatletter%
  \providecommand\color[2][]{%
    \errmessage{(Inkscape) Color is used for the text in Inkscape, but the package 'color.sty' is not loaded}%
    \renewcommand\color[2][]{}%
  }%
  \providecommand\transparent[1]{%
    \errmessage{(Inkscape) Transparency is used (non-zero) for the text in Inkscape, but the package 'transparent.sty' is not loaded}%
    \renewcommand\transparent[1]{}%
  }%
  \providecommand\rotatebox[2]{#2}%
  \newcommand*\fsize{\dimexpr\f@size pt\relax}%
  \newcommand*\lineheight[1]{\fontsize{\fsize}{#1\fsize}\selectfont}%
  \ifx\svgwidth\undefined%
    \setlength{\unitlength}{2678.99990845bp}%
    \ifx\svgscale\undefined%
      \relax%
    \else%
      \setlength{\unitlength}{\unitlength * \real{\svgscale}}%
    \fi%
  \else%
    \setlength{\unitlength}{\svgwidth}%
  \fi%
  \global\let\svgwidth\undefined%
  \global\let\svgscale\undefined%
  \makeatother%
  \begin{picture}(1,0.41955957)%
    \lineheight{1}%
    \setlength\tabcolsep{0pt}%
    \put(0,0){\includegraphics[width=\unitlength,page=1]{images/genz-pdf.pdf}}%
    \put(0.05861773,0.3967278){\makebox(0,0)[lt]{\lineheight{1.25}\smash{\begin{tabular}[t]{l}\small\textsc{in}\end{tabular}}}}%
    \put(0.93281434,0.3967278){\makebox(0,0)[lt]{\lineheight{1.25}\smash{\begin{tabular}[t]{l}\small\textsc{se}\end{tabular}}}}%
    \put(0.14453872,0.3967278){\makebox(0,0)[lt]{\lineheight{1.25}\smash{\begin{tabular}[t]{l}\small\textsc{dm}\end{tabular}}}}%
    \put(0.25134809,0.3967278){\makebox(0,0)[lt]{\lineheight{1.25}\smash{\begin{tabular}[t]{l}\small\textsc{sp}\end{tabular}}}}%
    \put(0.321405,0.3967278){\makebox(0,0)[lt]{\lineheight{1.25}\smash{\begin{tabular}[t]{l}\small\textsc{sp:ws}\end{tabular}}}}%
    \put(0.44770366,0.3967278){\makebox(0,0)[lt]{\lineheight{1.25}\smash{\begin{tabular}[t]{l}\small\textsc{se}\end{tabular}}}}%
    \put(0.54372838,0.3967278){\makebox(0,0)[lt]{\lineheight{1.25}\smash{\begin{tabular}[t]{l}\small\textsc{in}\end{tabular}}}}%
    \put(0.62964945,0.3967278){\makebox(0,0)[lt]{\lineheight{1.25}\smash{\begin{tabular}[t]{l}\small\textsc{dm}\end{tabular}}}}%
    \put(0.73645878,0.3967278){\makebox(0,0)[lt]{\lineheight{1.25}\smash{\begin{tabular}[t]{l}\small\textsc{sp}\end{tabular}}}}%
    \put(0.80651563,0.3967278){\makebox(0,0)[lt]{\lineheight{1.25}\smash{\begin{tabular}[t]{l}\small\textsc{sp:ws}\end{tabular}}}}%
    \put(0.02264152,0.24962636){\rotatebox{90}{\makebox(0,0)[lt]{\lineheight{1.25}\smash{\begin{tabular}[t]{l}\small\textsc{chairs}\end{tabular}}}}}%
    \put(0.02264152,0.06105719){\rotatebox{90}{\makebox(0,0)[lt]{\lineheight{1.25}\smash{\begin{tabular}[t]{l}\small\textsc{vases}\end{tabular}}}}}%
  \end{picture}%
\endgroup%

%% file: images/geometric-pdf.pdf_tex
\begingroup%
  \makeatletter%
  \providecommand\color[2][]{%
    \errmessage{(Inkscape) Color is used for the text in Inkscape, but the package 'color.sty' is not loaded}%
    \renewcommand\color[2][]{}%
  }%
  \providecommand\transparent[1]{%
    \errmessage{(Inkscape) Transparency is used (non-zero) for the text in Inkscape, but the package 'transparent.sty' is not loaded}%
    \renewcommand\transparent[1]{}%
  }%
  \providecommand\rotatebox[2]{#2}%
  \newcommand*\fsize{\dimexpr\f@size pt\relax}%
  \newcommand*\lineheight[1]{\fontsize{\fsize}{#1\fsize}\selectfont}%
  \ifx\svgwidth\undefined%
    \setlength{\unitlength}{2713.99987793bp}%
    \ifx\svgscale\undefined%
      \relax%
    \else%
      \setlength{\unitlength}{\unitlength * \real{\svgscale}}%
    \fi%
  \else%
    \setlength{\unitlength}{\svgwidth}%
  \fi%
  \global\let\svgwidth\undefined%
  \global\let\svgscale\undefined%
  \makeatother%
  \begin{picture}(1,0.31687548)%
    \lineheight{1}%
    \setlength\tabcolsep{0pt}%
    \put(0.0707579,0.30031941){\makebox(0,0)[lt]{\lineheight{1.25}\smash{\begin{tabular}[t]{l}\small\textsc{in}\end{tabular}}}}%
    \put(0.93368081,0.30031941){\makebox(0,0)[lt]{\lineheight{1.25}\smash{\begin{tabular}[t]{l}\small\textsc{se}\end{tabular}}}}%
    \put(0.15557085,0.30031941){\makebox(0,0)[lt]{\lineheight{1.25}\smash{\begin{tabular}[t]{l}\small\textsc{dm}\end{tabular}}}}%
    \put(0.26100279,0.30031941){\makebox(0,0)[lt]{\lineheight{1.25}\smash{\begin{tabular}[t]{l}\small\textsc{sp}\end{tabular}}}}%
    \put(0.33015625,0.30031941){\makebox(0,0)[lt]{\lineheight{1.25}\smash{\begin{tabular}[t]{l}\small\textsc{sp:ws}\end{tabular}}}}%
    \put(0.45482615,0.30031941){\makebox(0,0)[lt]{\lineheight{1.25}\smash{\begin{tabular}[t]{l}\small\textsc{se}\end{tabular}}}}%
    \put(0.54961253,0.30031941){\makebox(0,0)[lt]{\lineheight{1.25}\smash{\begin{tabular}[t]{l}\small\textsc{in}\end{tabular}}}}%
    \put(0.63442556,0.30031941){\makebox(0,0)[lt]{\lineheight{1.25}\smash{\begin{tabular}[t]{l}\small\textsc{dm}\end{tabular}}}}%
    \put(0.73985746,0.30031941){\makebox(0,0)[lt]{\lineheight{1.25}\smash{\begin{tabular}[t]{l}\small\textsc{sp}\end{tabular}}}}%
    \put(0.80901085,0.30031941){\makebox(0,0)[lt]{\lineheight{1.25}\smash{\begin{tabular}[t]{l}\small\textsc{sp:ws}\end{tabular}}}}%
    \put(0,0){\includegraphics[width=\unitlength,page=1]{images/geometric-pdf.pdf}}%
    \put(0.02057384,0.06406764){\rotatebox{90}{\makebox(0,0)[lt]{\lineheight{1.25}\smash{\begin{tabular}[t]{l}\small\textsc{geometric}\end{tabular}}}}}%
  \end{picture}%
\endgroup%

%% file: images/geo-to-org-pdf.pdf_tex
\begingroup%
  \makeatletter%
  \providecommand\color[2][]{%
    \errmessage{(Inkscape) Color is used for the text in Inkscape, but the package 'color.sty' is not loaded}%
    \renewcommand\color[2][]{}%
  }%
  \providecommand\transparent[1]{%
    \errmessage{(Inkscape) Transparency is used (non-zero) for the text in Inkscape, but the package 'transparent.sty' is not loaded}%
    \renewcommand\transparent[1]{}%
  }%
  \providecommand\rotatebox[2]{#2}%
  \newcommand*\fsize{\dimexpr\f@size pt\relax}%
  \newcommand*\lineheight[1]{\fontsize{\fsize}{#1\fsize}\selectfont}%
  \ifx\svgwidth\undefined%
    \setlength{\unitlength}{2715.99966431bp}%
    \ifx\svgscale\undefined%
      \relax%
    \else%
      \setlength{\unitlength}{\unitlength * \real{\svgscale}}%
    \fi%
  \else%
    \setlength{\unitlength}{\svgwidth}%
  \fi%
  \global\let\svgwidth\undefined%
  \global\let\svgscale\undefined%
  \makeatother%
  \begin{picture}(1,0.19293079)%
    \lineheight{1}%
    \setlength\tabcolsep{0pt}%
    \put(0,0){\includegraphics[width=\unitlength,page=1]{images/geo-to-org-pdf.pdf}}%
    \put(0.02129508,0.03860852){\rotatebox{90}{\makebox(0,0)[lt]{\lineheight{1.25}\smash{\begin{tabular}[t]{l}\small\textsc{similar}\end{tabular}}}}}%
    \put(0,0){\includegraphics[width=\unitlength,page=2]{images/geo-to-org-pdf.pdf}}%
  \end{picture}%
\endgroup%

%% file: images/organics-pdf.pdf_tex
\begingroup%
  \makeatletter%
  \providecommand\color[2][]{%
    \errmessage{(Inkscape) Color is used for the text in Inkscape, but the package 'color.sty' is not loaded}%
    \renewcommand\color[2][]{}%
  }%
  \providecommand\transparent[1]{%
    \errmessage{(Inkscape) Transparency is used (non-zero) for the text in Inkscape, but the package 'transparent.sty' is not loaded}%
    \renewcommand\transparent[1]{}%
  }%
  \providecommand\rotatebox[2]{#2}%
  \newcommand*\fsize{\dimexpr\f@size pt\relax}%
  \newcommand*\lineheight[1]{\fontsize{\fsize}{#1\fsize}\selectfont}%
  \ifx\svgwidth\undefined%
    \setlength{\unitlength}{2715.99993896bp}%
    \ifx\svgscale\undefined%
      \relax%
    \else%
      \setlength{\unitlength}{\unitlength * \real{\svgscale}}%
    \fi%
  \else%
    \setlength{\unitlength}{\svgwidth}%
  \fi%
  \global\let\svgwidth\undefined%
  \global\let\svgscale\undefined%
  \makeatother%
  \begin{picture}(1,0.57879237)%
    \lineheight{1}%
    \setlength\tabcolsep{0pt}%
    \put(0,0){\includegraphics[width=\unitlength,page=1]{images/organics-pdf.pdf}}%
    \put(0.02129517,0.21926141){\rotatebox{90}{\makebox(0,0)[lt]{\lineheight{1.25}\smash{\begin{tabular}[t]{l}\small\textsc{organics}\end{tabular}}}}}%
    \put(0,0){\includegraphics[width=\unitlength,page=2]{images/organics-pdf.pdf}}%
  \end{picture}%
\endgroup%

%% file: images/user-hand-pdf.pdf_tex
\begingroup%
  \makeatletter%
  \providecommand\color[2][]{%
    \errmessage{(Inkscape) Color is used for the text in Inkscape, but the package 'color.sty' is not loaded}%
    \renewcommand\color[2][]{}%
  }%
  \providecommand\transparent[1]{%
    \errmessage{(Inkscape) Transparency is used (non-zero) for the text in Inkscape, but the package 'transparent.sty' is not loaded}%
    \renewcommand\transparent[1]{}%
  }%
  \providecommand\rotatebox[2]{#2}%
  \newcommand*\fsize{\dimexpr\f@size pt\relax}%
  \newcommand*\lineheight[1]{\fontsize{\fsize}{#1\fsize}\selectfont}%
  \ifx\svgwidth\undefined%
    \setlength{\unitlength}{2173.00012207bp}%
    \ifx\svgscale\undefined%
      \relax%
    \else%
      \setlength{\unitlength}{\unitlength * \real{\svgscale}}%
    \fi%
  \else%
    \setlength{\unitlength}{\svgwidth}%
  \fi%
  \global\let\svgwidth\undefined%
  \global\let\svgscale\undefined%
  \makeatother%
  \begin{picture}(1,1.14091128)%
    \lineheight{1}%
    \setlength\tabcolsep{0pt}%
    \put(0,0){\includegraphics[width=\unitlength,page=1]{images/user-hand-pdf.pdf}}%
    \put(0.09159824,1.11710469){\makebox(0,0)[lt]{\lineheight{1.25}\smash{\begin{tabular}[t]{l}\small\textsc{ill}\end{tabular}}}}%
    \put(0.19977941,1.11710469){\makebox(0,0)[lt]{\lineheight{1.25}\smash{\begin{tabular}[t]{l}\small\textsc{cnt}\end{tabular}}}}%
    \put(0.32795482,1.11710469){\makebox(0,0)[lt]{\lineheight{1.25}\smash{\begin{tabular}[t]{l}\small\textsc{sk}\end{tabular}}}}%
    \put(0.5923122,1.11710469){\makebox(0,0)[lt]{\lineheight{1.25}\smash{\begin{tabular}[t]{l}\small\textsc{user sketches}\end{tabular}}}}%
    \put(0.01964887,0.8730376){\rotatebox{90}{\makebox(0,0)[lt]{\lineheight{1.25}\smash{\begin{tabular}[t]{l}\small\textsc{cubes}\end{tabular}}}}}%
    \put(0.02323462,0.46598744){\rotatebox{90}{\makebox(0,0)[lt]{\lineheight{1.25}\smash{\begin{tabular}[t]{l}\small\textsc{primitives}\end{tabular}}}}}%
    \put(0.01964887,0.09604939){\rotatebox{90}{\makebox(0,0)[lt]{\lineheight{1.25}\smash{\begin{tabular}[t]{l}\small\textsc{complexes}\end{tabular}}}}}%
  \end{picture}%
\endgroup%

%% file: egpaper.bbl
\begin{thebibliography}{10}\itemsep=-1pt

\bibitem{Benard:2013:SAE:2461912.2461929}
Pierre B{\'e}nard, Forrester Cole, Michael Kass, Igor Mordatch, James Hegarty,
  Martin~Sebastian Senn, Kurt Fleischer, Davide Pesare, and Katherine Breeden.
\newblock Stylizing {{Animation}} by {{Example}}.
\newblock {\em ACM Trans. Graph.}, 32(4):119:1--119:12, July 2013.

\bibitem{blender_online_community_blender_2019}
{Blender Online Community}.
\newblock {\em Blender - a {{3D}} Modelling and Rendering Package}.
\newblock {Blender Foundation}, {Blender Institute, Amsterdam}, 2019.

\bibitem{6654137}
C. Cao, Y. Weng, S. Zhou, Y. Tong, and K. Zhou.
\newblock {{FaceWarehouse}}: {{A 3D Facial Expression Database}} for {{Visual
  Computing}}.
\newblock {\em IEEE Transactions on Visualization and Computer Graphics},
  20(3):413--425, Mar. 2014.

\bibitem{shapenet2015}
Angel~X. Chang, Thomas Funkhouser, Leonidas Guibas, Pat Hanrahan, Qixing Huang,
  Zimo Li, Silvio Savarese, Manolis Savva, Shuran Song, Hao Su, Jianxiong Xiao,
  Li Yi, and Fisher Yu.
\newblock {{ShapeNet}}: {{An Information}}-{{Rich 3D Model Repository}}.
\newblock Technical Report arXiv:1512.03012 [cs.GR], {Stanford University ---
  Princeton University --- Toyota Technological Institute at Chicago}, 2015.

\bibitem{Chen_2018_CVPR}
Yang Chen, Yu-Kun Lai, and Yong-Jin Liu.
\newblock {{CartoonGAN}}: {{Generative Adversarial Networks}} for {{Photo
  Cartoonization}}.
\newblock In {\em The {{IEEE Conference}} on {{Computer Vision}} and {{Pattern
  Recognition}} ({{CVPR}})}, June 2018.

\bibitem{Curtis:1997:CW:258734.258896}
Cassidy~J. Curtis, Sean~E. Anderson, Joshua~E. Seims, Kurt~W. Fleischer, and
  David~H. Salesin.
\newblock Computer-generated {{Watercolor}}.
\newblock In {\em Proceedings of the 24th {{Annual Conference}} on {{Computer
  Graphics}} and {{Interactive Techniques}}}, {{SIGGRAPH}} '97, pages 421--430.
  {ACM Press/Addison-Wesley Publishing Co.}, 1997.

\bibitem{DeCarlo:2002:SAP:566654.566650}
Doug DeCarlo and Anthony Santella.
\newblock Stylization and {{Abstraction}} of {{Photographs}}.
\newblock {\em ACM Trans. Graph.}, 21(3):769--776, July 2002.

\bibitem{Delanoy:2018:SUM:3242771.3203197}
Johanna Delanoy, Mathieu Aubry, Phillip Isola, Alexei~A. Efros, and Adrien
  Bousseau.
\newblock {{3D Sketching Using Multi}}-{{View Deep Volumetric Prediction}}.
\newblock {\em Proc. ACM Comput. Graph. Interact. Tech.}, 1(1):21:1--21:22,
  July 2018.

\bibitem{Fan_2017_CVPR}
Haoqiang Fan, Hao Su, and Leonidas~J. Guibas.
\newblock A {{Point Set Generation Network}} for {{3D Object Reconstruction
  From}} a {{Single Image}}.
\newblock In {\em The {{IEEE Conference}} on {{Computer Vision}} and {{Pattern
  Recognition}} ({{CVPR}})}, July 2017.

\bibitem{fiser_stylit_2016}
Jakub Fi{\v s}er, Ond{\v r}ej Jamri{\v s}ka, Michal Luk{\'a}{\v c}, Eli
  Shechtman, Paul Asente, Jingwan Lu, and Daniel S{\'y}kora.
\newblock {{StyLit}}: Illumination-guided example-based stylization of {{3D}}
  renderings.
\newblock {\em ACM Trans. Graph.}, 35(4), July 2016.

\bibitem{Fiser:2017:ESS:3072959.3073660}
Jakub Fi{\v s}er, Ond{\v r}ej Jamri{\v s}ka, David Simons, Eli Shechtman,
  Jingwan Lu, Paul Asente, Michal Luk{\'a}{\v c}, and Daniel S{\'y}kora.
\newblock Example-based {{Synthesis}} of {{Stylized Facial Animations}}.
\newblock {\em ACM Trans. Graph.}, 36(4):155:1--155:11, July 2017.

\bibitem{furusawa_comicolorization_2017}
Chie Furusawa, Kazuyuki Hiroshiba, Keisuke Ogaki, and Yuri Odagiri.
\newblock Comicolorization: {{Semi}}-{{Automatic Manga Colorization}}.
\newblock {\em arXiv:1706.06759 [cs]}, Sept. 2017.

\bibitem{gerl:hal-00781065}
Moritz Gerl and Tobias Isenberg.
\newblock Interactive {{Example}}-based {{Hatching}}.
\newblock {\em Computers and Graphics}, 37(1-2):65--80, 2013.

\bibitem{NIPS2014_5423}
Ian Goodfellow, Jean {Pouget-Abadie}, Mehdi Mirza, Bing Xu, David
  {Warde-Farley}, Sherjil Ozair, Aaron Courville, and Yoshua Bengio.
\newblock Generative {{Adversarial Nets}}.
\newblock In Z. Ghahramani, M. Welling, C. Cortes, N.~D. Lawrence, and K.~Q.
  Weinberger, editors, {\em Advances in {{Neural Information Processing
  Systems}} 27}, pages 2672--2680. {Curran Associates, Inc.}, 2014.

\bibitem{Haeberli:1990:PNA:97880.97902}
Paul Haeberli.
\newblock Paint by {{Numbers}}: {{Abstract Image Representations}}.
\newblock {\em SIGGRAPH Comput. Graph.}, 24(4):207--214, Sept. 1990.

\bibitem{Han:2017:DDL:3072959.3073629}
Xiaoguang Han, Chang Gao, and Yizhou Yu.
\newblock {{DeepSketch2Face}}: {{A Deep Learning Based Sketching System}} for
  {{3D Face}} and {{Caricature Modeling}}.
\newblock {\em ACM Trans. Graph.}, 36(4):126:1--126:12, July 2017.

\bibitem{Hertzmann:1998:PRC:280814.280951}
Aaron Hertzmann.
\newblock Painterly {{Rendering}} with {{Curved Brush Strokes}} of {{Multiple
  Sizes}}.
\newblock In {\em Proceedings of the 25th {{Annual Conference}} on {{Computer
  Graphics}} and {{Interactive Techniques}}}, {{SIGGRAPH}} '98, pages 453--460.
  {ACM}, 1998.

\bibitem{Hertzmann:2001:IA:383259.383295}
Aaron Hertzmann, Charles~E. Jacobs, Nuria Oliver, Brian Curless, and David~H.
  Salesin.
\newblock Image {{Analogies}}.
\newblock In {\em Proceedings of the 28th {{Annual Conference}} on {{Computer
  Graphics}} and {{Interactive Techniques}}}, {{SIGGRAPH}} '01, pages 327--340.
  {ACM}, 2001.

\bibitem{Hertzmann:2000:ISS:344779.345074}
Aaron Hertzmann and Denis Zorin.
\newblock Illustrating {{Smooth Surfaces}}.
\newblock In {\em Proceedings of the 27th {{Annual Conference}} on {{Computer
  Graphics}} and {{Interactive Techniques}}}, {{SIGGRAPH}} '00, pages 517--526.
  {ACM Press/Addison-Wesley Publishing Co.}, 2000.

\bibitem{hu_squeeze-and-excitation_2018}
Jie Hu, Li Shen, and Gang Sun.
\newblock Squeeze-and-{{Excitation Networks}}.
\newblock In {\em Proceedings of the {{IEEE Conference}} on {{Computer Vision}}
  and {{Pattern Recognition}}}, pages 7132--7141, 2018.

\bibitem{Iarussi:2015:BRC:2774971.2710026}
Emmanuel Iarussi, David Bommes, and Adrien Bousseau.
\newblock {{BendFields}}: {{Regularized Curvature Fields}} from {{Rough Concept
  Sketches}}.
\newblock {\em ACM Trans. Graph.}, 34(3):24:1--24:16, May 2015.

\bibitem{Igarashi:2003:SMS:641480.641507}
Takeo Igarashi and John~F. Hughes.
\newblock Smooth {{Meshes}} for {{Sketch}}-based {{Freeform Modeling}}.
\newblock In {\em Proceedings of the 2003 {{Symposium}} on {{Interactive 3D
  Graphics}}}, {{I3D}} '03, pages 139--142. {ACM}, 2003.

\bibitem{Igarashi:1999:TSI:311535.311602}
Takeo Igarashi, Satoshi Matsuoka, and Hidehiko Tanaka.
\newblock Teddy: {{A Sketching Interface}} for {{3D Freeform Design}}.
\newblock In {\em Proceedings of the 26th {{Annual Conference}} on {{Computer
  Graphics}} and {{Interactive Techniques}}}, {{SIGGRAPH}} '99, pages 409--416.
  {ACM Press/Addison-Wesley Publishing Co.}, 1999.

\bibitem{isola_image--image_2016}
Phillip Isola, Jun-Yan Zhu, Tinghui Zhou, and Alexei~A. Efros.
\newblock Image-to-{{Image Translation}} with {{Conditional Adversarial
  Networks}}.
\newblock {\em arXiv:1611.07004 [cs]}, Nov. 2016.

\bibitem{Jamriska:2015:LAT:2809654.2766983}
Ond{\v r}ej Jamri{\v s}ka, Jakub Fi{\v s}er, Paul Asente, Jingwan Lu, Eli
  Shechtman, and Daniel S{\'y}kora.
\newblock {{LazyFluids}}: {{Appearance Transfer}} for {{Fluid Animations}}.
\newblock {\em ACM Trans. Graph.}, 34(4):92:1--92:10, July 2015.

\bibitem{942087}
D. Kang, D. Kim, and K. Yoon.
\newblock A study on the real-time toon rendering for {{3D}} geometry model.
\newblock In {\em Proceedings {{Fifth International Conference}} on
  {{Information Visualisation}}}, pages 391--396, July 2001.

\bibitem{kim_tag2pix_2019}
Hyunsu Kim, Ho~Young Jhoo, Eunhyeok Park, and Sungjoo Yoo.
\newblock {{Tag2Pix}}: {{Line Art Colorization Using Text Tag With SECat}} and
  {{Changing Loss}}.
\newblock In {\em Proceedings of the {{IEEE}}/{{CVF International Conference}}
  on {{Computer Vision}}}, pages 9056--9065, 2019.

\bibitem{krizhevsky_imagenet_2012}
Alex Krizhevsky, Ilya Sutskever, and Geoffrey~E. Hinton.
\newblock {{ImageNet}} classification with deep convolutional neural networks.
\newblock In {\em Proceedings of the 25th {{International Conference}} on
  {{Neural Information Processing Systems}} - {{Volume}} 1}, {{NIPS}}'12, pages
  1097--1105, {Red Hook, NY, USA}, Dec. 2012. {Curran Associates Inc.}

\bibitem{li_im2pencil_2019}
Yijun Li, Chen Fang, Aaron Hertzmann, Eli Shechtman, and Ming-Hsuan Yang.
\newblock {{Im2Pencil}}: {{Controllable Pencil Illustration From Photographs}}.
\newblock In {\em Proceedings of the {{IEEE}}/{{CVF Conference}} on {{Computer
  Vision}} and {{Pattern Recognition}}}, pages 1525--1534, 2019.

\bibitem{LIPSON1996651}
H Lipson and M Shpitalni.
\newblock Optimization-based reconstruction of a {{3D}} object from a single
  freehand line drawing.
\newblock {\em Computer-Aided Design}, 28(8):651--663, 1996.

\bibitem{Litwinowicz:1997:PIV:258734.258893}
Peter Litwinowicz.
\newblock Processing {{Images}} and {{Video}} for an {{Impressionist Effect}}.
\newblock In {\em Proceedings of the 24th {{Annual Conference}} on {{Computer
  Graphics}} and {{Interactive Techniques}}}, {{SIGGRAPH}} '97, pages 407--414.
  {ACM Press/Addison-Wesley Publishing Co.}, 1997.

\bibitem{long_fully_2015}
Jonathan Long, Evan Shelhamer, and Trevor Darrell.
\newblock Fully {{Convolutional Networks}} for {{Semantic Segmentation}}.
\newblock In {\em The {{IEEE Conference}} on {{Computer Vision}} and {{Pattern
  Recognition}} ({{CVPR}})}, June 2015.

\bibitem{Mackinlay:1988:DEP:1402242.1402254}
Jock Mackinlay, Steven Feiner, Jim Blinn, Donald~P. Greenberg, and Margaret~A.
  Hagen.
\newblock Designing {{Effective Pictures}}: {{Is Photographic Realism}} the
  {{Only Answer}}?
\newblock In {\em {{ACM SIGGRAPH}} 88 {{Panel Proceedings}}}, {{SIGGRAPH}} '88,
  pages 12:1--12:48. {ACM}, 1988.

\bibitem{24791}
J. Malik and D. Maydan.
\newblock Recovering three-dimensional shape from a single image of curved
  objects.
\newblock {\em IEEE Transactions on Pattern Analysis and Machine Intelligence},
  11(6):555--566, June 1989.

\bibitem{DBLP:journals/corr/MirzaO14}
Mehdi Mirza and Simon Osindero.
\newblock Conditional {{Generative Adversarial Nets}}.
\newblock {\em CoRR}, abs/1411.1784, 2014.

\bibitem{Mori:2007:PID:1276377.1276433}
Yuki Mori and Takeo Igarashi.
\newblock Plushie: {{An Interactive Design System}} for {{Plush Toys}}.
\newblock {\em ACM Trans. Graph.}, 26(3), July 2007.

\bibitem{praun_real-time_2001}
Emil Praun, Hugues Hoppe, Matthew Webb, and Adam Finkelstein.
\newblock Real-time hatching.
\newblock In {\em Proceedings of the 28th Annual Conference on {{Computer}}
  Graphics and Interactive Techniques - {{SIGGRAPH}} '01}, page 581, {Not
  Known}, 2001. {ACM Press}.

\bibitem{ronneberger_u-net_2015}
Olaf Ronneberger, Philipp Fischer, and Thomas Brox.
\newblock U-{{Net}}: {{Convolutional Networks}} for {{Biomedical Image
  Segmentation}}.
\newblock {\em arXiv:1505.04597 [cs]}, May 2015.

\bibitem{salimans2016improved}
Tim Salimans, Ian Goodfellow, Wojciech Zaremba, Vicki Cheung, Alec Radford, Xi
  Chen, and Xi Chen.
\newblock Improved {{Techniques}} for {{Training GANs}}.
\newblock In D.~D. Lee, M. Sugiyama, U.~V. Luxburg, I. Guyon, and R. Garnett,
  editors, {\em Advances in {{Neural Information Processing Systems}} 29},
  pages 2234--2242. {Curran Associates, Inc.}, 2016.

\bibitem{Salisbury:1994:IPI:192161.192185}
Michael~P. Salisbury, Sean~E. Anderson, Ronen Barzel, and David~H. Salesin.
\newblock Interactive {{Pen}}-and-ink {{Illustration}}.
\newblock In {\em Proceedings of the 21st {{Annual Conference}} on {{Computer
  Graphics}} and {{Interactive Techniques}}}, {{SIGGRAPH}} '94, pages 101--108.
  {ACM}, 1994.

\bibitem{Santella:2002:APR:508530.508544}
Anthony Santella and Doug DeCarlo.
\newblock Abstracted {{Painterly Renderings Using Eye}}-tracking {{Data}}.
\newblock In {\em Proceedings of the 2nd {{International Symposium}} on
  {{Non}}-Photorealistic {{Animation}} and {{Rendering}}}, {{NPAR}} '02, pages
  75--ff. {ACM}, 2002.

\bibitem{shao:hal-00703202}
Cloud Shao, Adrien Bousseau, Alla Sheffer, and Karan Singh.
\newblock {{CrossShade}}: {{Shading Concept Sketches Using Cross}}-{{Section
  Curves}}.
\newblock {\em ACM Transactions on Graphics}, 31(4), 2012.

\bibitem{Shiraishi:2000:AAP:340916.340923}
Michio Shiraishi and Yasushi Yamaguchi.
\newblock An {{Algorithm}} for {{Automatic Painterly Rendering Based}} on
  {{Local Source Image Approximation}}.
\newblock In {\em Proceedings of the 1st {{International Symposium}} on
  {{Non}}-Photorealistic {{Animation}} and {{Rendering}}}, {{NPAR}} '00, pages
  53--58. {ACM}, 2000.

\bibitem{simo-serra_mastering_2018}
Edgar {Simo-Serra}, Satoshi Iizuka, and Hiroshi Ishikawa.
\newblock Mastering {{Sketching}}: {{Adversarial Augmentation}} for
  {{Structured Prediction}}.
\newblock {\em ACM Transactions on Graphics}, 37(1):11:1--11:13, Jan. 2018.

\bibitem{simo-serra_real-time_2018}
Edgar {Simo-Serra}, Satoshi Iizuka, and Hiroshi Ishikawa.
\newblock Real-time data-driven interactive rough sketch inking.
\newblock {\em ACM Transactions on Graphics}, 37(4):98:1--98:14, July 2018.

\bibitem{simo-serra_learning_2016}
Edgar {Simo-Serra}, Satoshi Iizuka, Kazuma Sasaki, and Hiroshi Ishikawa.
\newblock Learning to simplify: Fully convolutional networks for rough sketch
  cleanup.
\newblock {\em ACM Transactions on Graphics}, 35(4):121:1--121:11, July 2016.

\bibitem{the_blender_foundation_freestyle_2019}
{The Blender Foundation}.
\newblock Freestyle \textemdash{} {{Blender Manual}}.
\newblock https://docs.blender.org/manual/en/2.81/render/freestyle/index.html,
  Dec-2019.

\bibitem{wang_high-resolution_2017}
Ting-Chun Wang, Ming-Yu Liu, Jun-Yan Zhu, Andrew Tao, Jan Kautz, and Bryan
  Catanzaro.
\newblock High-{{Resolution Image Synthesis}} and {{Semantic Manipulation}}
  with {{Conditional GANs}}.
\newblock {\em arXiv:1711.11585 [cs]}, Nov. 2017.

\bibitem{Xu:2015:ITS:2810210.2810212}
Q. Xu, Y. Gingold, and K. Singh.
\newblock Inverse {{Toon Shading}}: {{Interactive Normal Field Modeling}} with
  {{Isophotes}}.
\newblock In {\em Proceedings of the {{Workshop}} on {{Sketch}}-{{Based
  Interfaces}} and {{Modeling}}}, {{SBIM}} '15, pages 15--25. {Eurographics
  Association}, 2015.

\bibitem{Zeleznik:1996:SIS:237170.237238}
Robert~C. Zeleznik, Kenneth~P. Herndon, and John~F. Hughes.
\newblock {{SKETCH}}: {{An Interface}} for {{Sketching 3D Scenes}}.
\newblock In {\em Proceedings of the 23rd {{Annual Conference}} on {{Computer
  Graphics}} and {{Interactive Techniques}}}, {{SIGGRAPH}} '96, pages 163--170.
  {ACM}, 1996.

\bibitem{Zeng:2009:IPP:1640443.1640445}
Kun Zeng, Mingtian Zhao, Caiming Xiong, and Song-Chun Zhu.
\newblock From {{Image Parsing}} to {{Painterly Rendering}}.
\newblock {\em ACM Trans. Graph.}, 29(1):2:1--2:11, Dec. 2009.

\bibitem{zhang_two-stage_2018}
Lvmin Zhang, Chengze Li, Tien-Tsin Wong, Yi Ji, and Chunping Liu.
\newblock Two-stage sketch colorization.
\newblock {\em ACM Transactions on Graphics}, 37(6):261:1--261:14, Dec. 2018.

\bibitem{zheng_learning_2020}
Qingyuan Zheng, Zhuoru Li, and Adam Bargteil.
\newblock Learning to {{Shadow Hand}}-drawn {{Sketches}}.
\newblock In {\em {{CVPR}}}, page~28, 2020.

\bibitem{177}
Youyi Zheng, Han Liu, Julie Dorsey, and Niloy Mitra.
\newblock {{SMART CANVAS}} : {{Context}}-inferred {{Interpretation}} of
  {{Sketches}} for {{Preparatory Design Studies}}.
\newblock May 2016.

\end{thebibliography}
